\documentclass[sigconf,balance=false,authorversion]{acmart}
\usepackage{submission}

\usepackage{acmart-taps}

\usepackage{macros}
\usepackage{colortbl}
\usepackage{makecell}
\usepackage[bottom]{footmisc}
\usepackage{subcaption}
\usepackage{cleveref}
\usepackage[justification=centering]{caption}
\usepackage{xcolor}

\copyrightyear{2024}
\acmYear{2024}
\setcopyright{acmlicensed}\acmConference[CHI '24]{Proceedings of the CHI
Conference on Human Factors in Computing Systems}{May 11--16,
2024}{Honolulu, HI, USA}
\acmBooktitle{Proceedings of the CHI Conference on Human Factors in
Computing Systems (CHI '24), May 11--16, 2024, Honolulu, HI, USA}
\acmDOI{10.1145/3613904.3642566}
\acmISBN{979-8-4007-0330-0/24/05}

\begin{document}

\title{Understanding the Challenges of OpenSCAD Users for~3D~Printing}

\author{J. Felipe Gonzalez}
\orcid{0000-0002-0716-1689}
\affiliation{%
  \institution{Carleton University}
  \institution{Univ. Lille, CNRS, Inria, Centrale Lille, UMR 9189 CRIStAL}
  \postcode{F-59650}
  \city{Lille}
  \country{France}
}
\email{johannavila@cmail.carleton.ca}

\author{Thomas Pietrzak}
\orcid{0000-0002-2013-7253}
\affiliation{%
  \institution{Univ. Lille, CNRS, Inria, Centrale Lille, UMR 9189 CRIStAL}
  \postcode{F-59000}
  \city{Lille}
  \country{France}
}
\email{thomas.pietrzak@univ-lille.fr}

\author{Audrey Girouard}
\orcid{0000-0003-3223-105X}
\affiliation{%
  \institution{Carleton University}
  \city{Ottawa}
  \state{ON}
  \country{Canada}
}
\email{audrey.girouard@carleton.ca}

\author{G\'ery Casiez}
\orcid{0000-0003-1905-815X}
\affiliation{%
  \institution{Univ. Lille, CNRS, Inria, Centrale Lille, UMR 9189 CRIStAL}
  \postcode{F-59000}
  \city{Lille}
  \country{France}}
\additionalaffiliation{%
  \institution{Institut Universitaire de France}
  \city{Paris}
  \country{France}}
\email{gery.casiez@univ-lille.fr}
\renewcommand{\shortauthors}{J. Felipe Gonzalez, \textit{et al.}}

\begin{abstract}

Direct manipulation has been established as the main interaction paradigm for Computer-Aided Design (CAD) for decades.
It provides fast, incremental, and reversible actions that allow for an iterative process on a visual representation of the result.
Despite its numerous advantages, some users prefer a programming-based approach where they describe the 3D model they design with a specific programming language, such as OpenSCAD.
It allows users to create complex structured geometries and facilitates abstraction.
Unfortunately, most current knowledge about CAD practices only focuses on direct manipulation programs.
In this study, we interviewed 20 programming-based CAD users to understand their motivations and challenges.
Our findings reveal that this programming-oriented population presents difficulties in the design process in tasks such as 3D spatial understanding, validation and code debugging, creation of organic shapes, and code-view navigation.

\end{abstract}



\maketitle

\section{Introduction}

\textit{Computer-Aided Design} (CAD) applications are used to aid design processes across various fields, including the rapidly growing 3D printing community.
3D printing technology allows individuals to design and fabricate objects easily and quickly in what is known as the digital personal fabrication practice~\cite{ashbrook_towards_2016,mota_rise_2011}.
After identifying an idea or a need, makers create a digital model from scratch or retrieve it from model-storing websites~\cite{thingiversecom_thingiverse_2022}, edit it, and print it in hours.
Generally, users follow an iterative process going back and forth between the creativity, design, and printing stages until they achieve a satisfactory result~\cite{hudson_understanding_2016}.

Makers design using CAD applications that come in several flavors, integrate different technologies, and support different features.
Most of the available applications, such as TinkerCAD \cite{autodesk_tinkercad_2020}, FreeCAD \cite{the_freecad_team_freecad_2022}, or Fusion360 \cite{autodesk_inc_fusion_2023} use the \textbf{direct manipulation} interaction paradigm~\cite{shneiderman_direct_1983} where users can edit the models by making changes directly to their visual representation through simple metaphors such as drag-and-drop, menus, and buttons.
Direct manipulation provides immediate feedback, incremental and reversible operations~\cite{shneiderman_direct_1983,shneiderman_future_1982}, which enable a rapid learning curve~\cite{sherugar_direct_2016}.

A less popular category of CAD software use a \textbf{programming-based} approach that allows users to create 3D models by coding in a text editor with a specific programming language.
Users must create scripts describing the models, and the system compiles and renders the result in a 3D viewer.     Programming brings valuable advantages to 3D design.
For instance, repetitive actions can be easily generalized, such as placing multiple elements in a specified pattern, through the use of programmatic structures like conditionals and loops.
Moreover, programming allows the creation of very complex structured geometries, such as fractals or trees, through programming techniques like recursion.
It facilitates the utilization of mathematical formulas, version control, and abstraction~\cite{yeh_craftml_2018}.
Our primary focus lies in CAD applications that embrace the programming-based approach to its fullest extent.
Although some direct manipulation applications allow users to make modifications through code, as seen in FreeCAD with Python scripts \cite{freecad_python_2023}, this paper excludes such cases from the category of programming-based CAD because the code primarily executes specific actions rather than serving as a comprehensive model description.
In programming-based CAD, all model changes are reflected in the code, which always represents the complete model description. When modifications occur, the system re-executes all scripts to generate a new 3D model.
Open-SCAD~\cite{openscad_openscad_2020}, CadQuery~\cite{cadquery_cadquery_2023}, IceSL~\cite{lefebvre_icesl_2022}, and JSCad~\cite{openjscadorg_jscad_2023} are examples of programming-based CAD software.

Programming-based CAD applications are important players in CAD design and, specifically in the personal fabrication field.
They present a solution to design complex structured geometries and offer a different paradigm from direct manipulation, allowing an alternative way of 3D designing~\cite{chytas_learning_2018,kastl_3d_2017}.
However, the potential of programming-based CAD may be underestimated and its challenges may not have been adequately studied.
Previous research on user behavior in 3D printing design has focused on investigating improvement opportunities for CAD almost exclusively in direct manipulation applications~\cite{twigg-smith_tools_2021,kim_understanding_2017,mahapatra_barriers_2019,kolaric_comprehending_2010}, with findings specific to this context.
For instance, Mahapatra~\textit{et al.}~\cite{mahapatra_barriers_2019} studied the barriers to taking measurements of objects, transferring them to CAD applications, and manipulating this information digitally.
Some of the difficulties classified as \textit{`` Digital''} are specific to direct manipulation applications such as \textit{``3D camera causes manipulation errors''}.
Furthermore, the literature related to programming-based CAD applications is limited.
Previous research has examined the limitations of understanding and navigating code in programming-based CAD \cite{gonzalez_introducing_2023} and its potential for students to learn programming through 3D design~\cite{kastl_3d_2017}.
However, these studies address specific task difficulties within a limited scope of the 3D printing design experience.
A better understanding of the user experience with these applications, their advantages, and the problems users face when 3D printing is still missing.
We have set out to investigate the following research questions:
What are the motivations and challenges of using programming-based CAD?
What are the current limitations of these applications?
In the context of 3D printing, how do the challenges previously identified in direct manipulation CAD applications relate to the ones present in programming-based CAD applications?
There is an interesting opportunity for HCI to investigate the current challenges programming-based CAD users face to facilitate the design process and possibly lower the entry barrier for newcomers.

This paper investigates how users of programming-based CAD software experience the design and fabrication process.
We conducted semi-structured interviews with 20 users of the most popular programming-based CAD software in 3D printing for personal fabrication, OpenSCAD \cite{machado_parametric_2019,nilsiam_free_2017}.
We asked participants about their design experiences with programming-based CAD, direct manipulation applications, and comparisons between them in the design and printing process.
Additionally, we draw on previous work to explore their motivations and contrast barriers found in 3D printing with direct manipulation applications.
Specifically, we included questions related to problems measuring physical objects to create digital designs \cite{mahapatra_barriers_2019} and limitations working with pre-existing models from websites \cite{alcock_barriers_2016}.
We also included a short hands-on exercise to observe design workflows and difficulties \cite{yeh_craftml_2018,gonzalez_introducing_2023}.
Based on the findings of the interviews, we provide a comprehensive analysis of the preferences of programming-based CAD users, current challenges in the design and printing process, and desired features expressed by the participants to improve these applications.

Our contribution is a qualitative analysis that aims to provide a comprehensive understanding of the programming-based CAD population. Specifically, we examine their motivations, design challenges, and challenges in the application field of 3D printing. 
Our findings suggest that programming-based CAD users, often with a long programming experience and a programming-oriented mindset, face significant difficulties measuring and designing organic and curve shapes, mentally connecting the code with the view, performing spatial transformation due to the required mathematical skills, and addressing uncertainty when 3D printing.

\section{Background}\label{sec:background}

We describe some technological aspects of CAD software to frame our findings.
We begin by defining what a programming-based application is.
Then, we differentiate between the workflows of CAD applications, specifically parametric and direct modeling. Additionally, we explain the two primary data representations that CAD applications use. Finally, we introduce the CAD application we used for this study, OpenSCAD.

\subsection{Programming-based CAD}

\textit{Programming-based CAD} refers to the applications that allow users to describe models entirely through coded instructions while the system renders the result in a view.
In particular, the code represents the full description of the model, and any edit of the model is described in the code.
Text-based applications such as JSCad~\cite{openjscadorg_jscad_2023},
BRL-CAD~\cite{devcom_analysis_center_brl-cad_2023}, and OpenSCAD~\cite{openscad_openscad_2020} fall into this definition. CAD applications using visual programming such as BlockSCAD~\cite{blockscad_inc_blockscad_2023} or Grasshopper \cite{davidson_grasshopper_2023} can also be considered as programming-based CAD.

Some direct manipulation applications allow users to modify the model with code.
McGuffin and Fuhrman~\cite{mcguffin_categories_2020} present a taxonomy of applications using different interaction paradigms, which can be applied to CAD.
\textit{Content-Oriented Programming} paradigm enables the output to be affected by instructions and direct manipulation.
Blender~\cite{foundation_blenderorg_2023} is an example of such an application, with which users create and edit meshes in the view and run scripts for specific actions with a code editor.
Another category that uses both direct manipulation and coded instructions is \textit{Programming By Example}, where direct manipulation interactions generate instructions that the user can execute later to edit the model.
For instance, FreeCAD~\cite{freecad_python_2023} provides a console where actions in the view generate corresponding Python code statements that the user can run later.
These approaches present a fundamental workflow difference compared to programming-based applications.
The code is used to perform specific actions, but it does not comprehensively describe the 3D model visually represented.
In programming-based, a modification requires the user to go through the code, edit it coherently, and re-execute all the scripts to re-generate the model; in the other paradigms, the code executes specific actions to modify the current state of the model.

McGuffin and Fuhrman also present the \textit{Bidirectional Programming} applications, where users can modify the output with both direct manipulation and instructions.
In these applications, every change in the output through direct manipulation results in an update in the code to keep coherence, such as presented in the scalable vector graphics (SVG) environments Sketch-N-Sketch \cite{hempel_sketch-n-sketch_2019} or Twoville \cite{johnson_computational_2023}.
Bidirectional programming CAD applications \cite{keeter_antimony_2023,keeter_libfive_2022,gonzalez_introducing_2023} follow the definition of the programming-based CAD paradigm by always keeping the code as a full description of the model, but they also extend the interaction capability by allowing direct manipulation interactions in the view.
Antimony \cite{keeter_antimony_2023} is an example of a programming-based CAD that allows users to use visual programming to create and edit the model in the view while the system updates the instructions coherently.
Despite the potential benefits of these applications, they seldom seem to be used, given the difficulty of finding models online.

\subsection{Parametric and direct modeling}\label{subsec:parametric}

CAD applications can be divided into two groups based on how the system stores and executes changes in the model~\cite{zou_parametricdirect_2022}: parametric modeling and direct modeling.
Parametric modeling, also called feature-based or history-based modeling \cite{anderson_geometric_1990, kim_incremental_1993}, allows the user to describe the model in a re-executable set of steps defined by parameters.
Parameters are adjustable to create new versions of the model by re-executing the steps.
Programming-based CAD is naturally parametric due to the way coding works, having arguments as input that determine the output.
On the other hand, direct manipulation applications that apply a parametric approach commonly allow users to define constraints and perform operations to the model, also called features, stored in a \textit{history tree}.
Modifiable parameters often control these features.
Thus, besides verifying the history of the design process, the user can perform edits on the history tree and re-execute it.
Consequently, users can obtain solid model variants by editing parameters embedded in the model \cite{zou_parametricdirect_2022}.
FreeCAD \cite{the_freecad_team_freecad_2022} is an example of parametric modeling.
In contrast, direct modeling allows users to edit the model without worrying about the history of these edits.
The system only captures the current state of the model.
Consequently, users gain flexibility, such as not worrying about features and the impact changes may have on their inter-dependencies \cite{brunelli_parametric_2022}, by renouncing to revisit their steps~\cite{brunelli_parametric_2022}.
Tinkercad~\cite{amabilis_amabilis_2022} is an example of a direct modeling application.

\subsection{Data representation}

\begin{figure*}
\centering
\subcaptionbox[Caption for comparison between FreeCAD and OpenSCAD]{
FreeCAD is a parametric direct manipulation software using B-rep. (1) Users can check the history tree to revisit their steps. (2) They can also interact in the view, individualizing vertices, edges, or faces. (3) FreeCad provides a Python console in which users can execute specific actions through code.
\label{fig:FreeCAD}}{\includegraphics[width=0.47\textwidth]{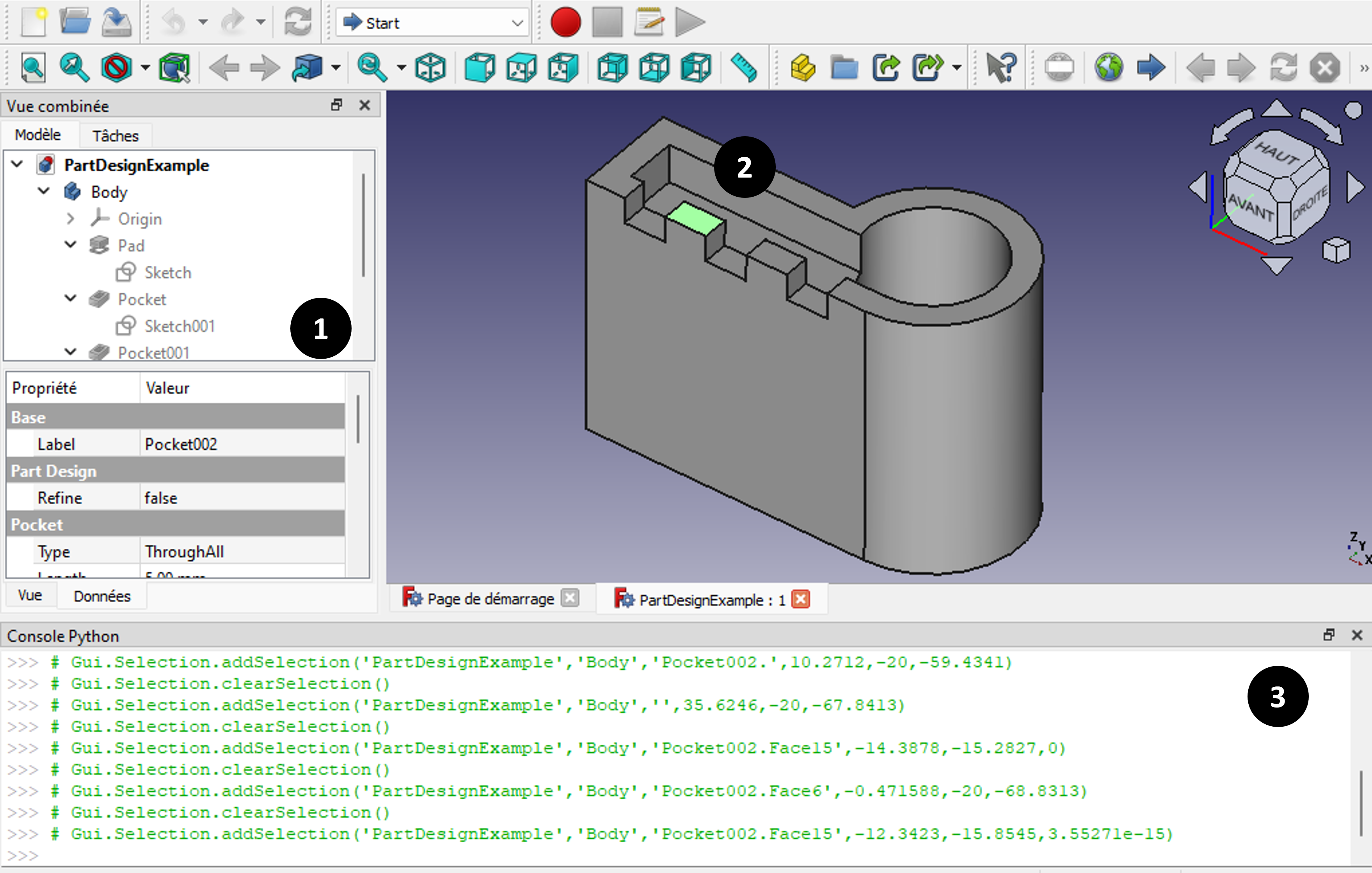}
}
\hspace{0.02\textwidth}
\subcaptionbox{OpenSCAD is parametric programming-based CAD software using CSG. (1) Users can describe the models through primitives (\textit{e.g.} sphere), transformations (\textit{e.g.} translate), and boolean operations (\textit{e.g.} union) in a text editor. (2) The system compiles and renders the result in a 3D viewer.
\label{fig:OpenSCAD}}{\includegraphics[width=0.47\textwidth]{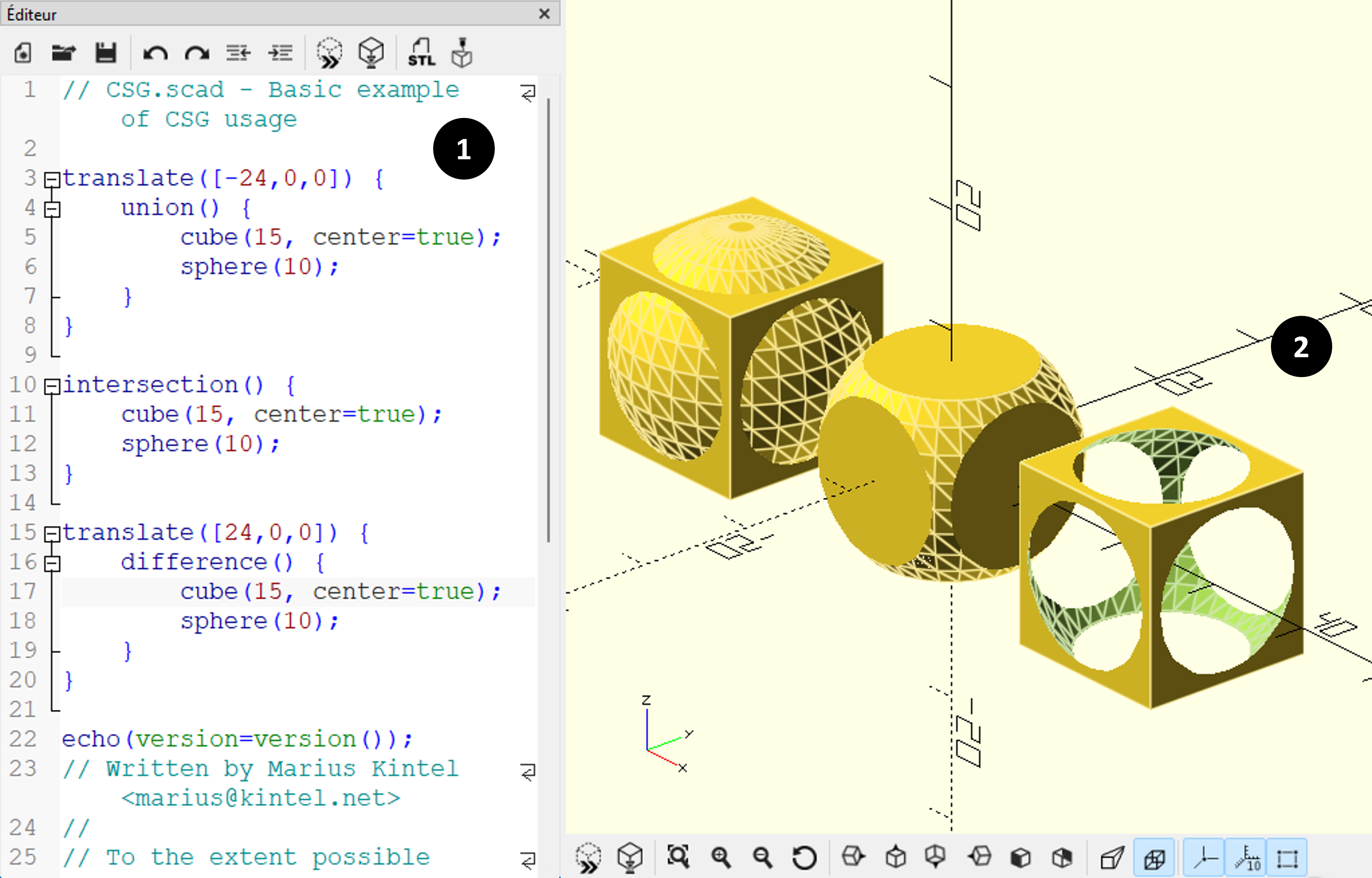}
}
\Description{Example of two parametric CAD programs. On the left, FreeCAD. FreeCAD is a parametric direct manipulation software using B-rep. Users can check the history tree to revisit their steps. They also can interact in the view, individualizing vertices, edges, or faces. FreeCad provides a Python console where users can execute specific actions through code. On the right, OpenSCAD. OpenSCAD is a parametric programming-based CAD software using CSG. Users can describe the models through primitives (e.g. sphere), transformations (e.g. translate), and boolean operations (e.g. union) in a text editor. The system compiles and renders the result in a 3D viewer.
}
\caption{Example of two parametric CAD applications.}
\label{figCADExamples}
\end{figure*}

Another distinguishing aspect of CAD applications is how the geometric data is represented and stored.
We distinguish between programs using constructive solid geometry (CSG) and boundary representation (B-rep).
In CSG, a solid is represented as a set of primitive solid objects (\textit{e.g.} spheres and cubes), transformations (\textit{e.g.} scale or mirror), and boolean operations (\textit{e.g.} union or intersection) as depicted in Figure \ref{fig:OpenSCAD}.
Both the surface and the interior of an object are implicitly defined.
In other words, there is no description of specific geometric properties, such as points, edges, or positions, but abstract descriptions of primitives and operations.
CSG objects are always \textit{watertight} and manifold \cite{stutz_formal_2018} if the primitives are \cite{hoffmann_geometric_1989}.
Therefore, CSG provides closed and well-formed geometries that are printable, making CSG attractive for 3D printing \cite{gibson_additive_2015}.
On the other hand, a boundary representation (B-rep) describes only the oriented surface of a solid as a data structure composed of vertices, edges, and faces.
B-rep can be efficiently rendered on a graphic display system, allowing easy differentiation between the vertices, edges, and faces (Figure \ref{fig:FreeCAD}).
This offers more flexibility than CSG

 and allows for valuable operations in 3D printing, such as chamfering, blending, or drafting ~\cite{gibson_additive_2015,hoffmann_geometric_1989}.
OpenSCAD~\cite{openscad_openscad_2020} uses a CSG representation while FreeCAD \cite{the_freecad_team_freecad_2022} uses a B-rep.

\subsection{OpenSCAD}\label{sec::os}

OpenSCAD is an open-source parametric CSG programming-based CAD application.
Users can describe 3D models in its \textit{functional declarative} programming language using the text editor, and the system compiles and renders the scripts in a viewer.
Although OpenSCAD has applications in various domains \cite{sleight_lasercutscad_2023}, it provides mainly 3D printing-oriented features.
For instance, OpenSCAD preferences menu offers features such as connecting with OctoPrint \cite{hausge_octoprintorg_2023}, a web interface for controlling consumer 3D printers or export models into the standard STL format for 3D printing \cite{library_of_congress_stl_2019}.
OpenSCAD aims to give full control over the design by being purely programming-based.
It is the most popular of the programming-based CAD applications, which are mainly parametric CSG applications such as IceSL~\cite{lefebvre_icesl_2022}, JSCad~\cite{openjscadorg_jscad_2023},
BRL-CAD~\cite{devcom_analysis_center_brl-cad_2023},  ImplicitCAD~\cite{longtin_implicitcadorg_2023}, or RapCAD~\cite{bathgate_rapcad_2023}.

OpenSCAD is not an interactive modeler and does not focus on the artistic aspects of 3D modeling but on the CAD aspects \cite{openscad_openscad_2020}.
In addition to CSG modeling techniques, it allows the extrusion of 2D outlines.
It also provides a \textit{preview} mode that generates approximations for rapid visualization and a \textit{render} mode that generates exact geometries using longer rendering times.
In preview mode, OpenSCAD allows the user to right-click on the models in the view to display a menu of the CSG elements that create the clicked part.
The user can click on a menu item while the system places the text cursor in the line of code that creates the CSG element.
Moreover, OpenSCAD language includes \textit{modifiers} for debugging.
Modifiers are specific characters that can be placed at the beginning of code statements to ignore, highlight, or isolate elements in the view\footnote{https://en.wikibooks.org/wiki/OpenSCAD\_User\_Manual/Modifier\_Characters. Accessed: 11/12/2023 }.
Finally, OpenSCAD handles command-line arguments and is not limited to the graphic user interface.

\section{Related work}

Creating 3D models is an essential part of the 3D printing process.
Makers can create a model from scratch using specialized CAD software.
Furthermore, they can get a pre-existing model from model-storing websites to print it as is or edit it in a CAD application to get a customized version.
We describe previous work investigating end-user behaviors and challenges in 3D design and 3D printing in such scenarios.
We start by looking at lessons learned from work on direct manipulation programs, and we continue with related work on programming-based CAD.
Finally, we describe some existing problems with model-storing websites.

\subsection{Modeling with direct manipulation programs}

The direct manipulation paradigm facilitates user interaction by reducing the cognitive resources required to understand and use user interfaces \cite{shneiderman_direct_1997, aish_designscript_2012}.
The actions must be fast, incremental, and reversible, while the objects of interest must be visible and directly manipulable~\cite{shneiderman_direct_1983,shneiderman_future_1982}.
Hence, this paradigm guarantees several usability benefits, such as recovering from errors and learning how to use the interface~\cite{sherugar_direct_2016}.
However, direct manipulation presents well-known challenges, such as difficulties in performing repetitive actions, manipulating small objects (especially in high-density spaces of objects), or manipulating intangible properties (abstract properties without visual form) \cite{frohlich_direct_1997,kwon_direct_2011}.

In 3D printing CAD, Hudson \textit{et al.}~\cite{hudson_understanding_2016} studied novices interacting with TinkerCAD and reported that manipulating elements in a 3D space through a 2D screen can confuse and lead to errors that are known problems in other digital applications~\cite{foley_computer_1996}.
Similar difficulties related to spatial thinking skills have also been reported in children using TinkerCAD~\cite{bhaduri_3dnst_2021}.
Specifically, problems related to understanding the 3D perspective, understanding rotation in a 3D space, using the correct primitive shapes, and grouping primitive shapes.
Programming-based CAD applications can present similar problems by having a 3D viewer rendering the output.

An everyday use of 3D printing is to repair \cite{hudson_understanding_2016} or to augment objects~\cite{ashbrook_towards_2016}.
In such cases, measurements of physical objects, their transfer, and their meaningful use in the design are necessary.
Mahapatra \textit{et al.}~\cite{mahapatra_barriers_2019} study how self-identified novice users capture physical measurements, transfer them to digital design, and verify their accuracy.
The study carried out on TinkerCAD reported several obstacles when novices create a digital replica of an object, classified into three groups according to the moment they occur: 1) \textit{Physical} when capturing the data, 2) \textit{Digital} when using the data in the digital design, and 3) \textit{Transition} when transferring (from physical to digital) or evaluating (from digital to physical) the data.
Although physical obstacles are independent of the modeling CAD, digital and transition may differ when using a programming-based application.
Probably some challenges may persist (\textit{e.g.} \textit{"3D camera causes confusion"}), others may occur differently (\textit{e.g.} \textit{"Relative placement problems"}), and others may not make sense in programming-based CAD (\textit{e.g.} \textit{"Miscalculating by hand"}).
We take inspiration from this work to explore what problems programming-based CAD users face when measuring, transferring, and using data in the design process.

\subsection{Modeling with programming-based CAD}

Significant efforts have been made to understand the difficulties in learning programming languages \cite{ko_six_2004,robins_learning_2003}.
Expectedly, programming-based CAD users may present similar challenges, such as difficulties in structuring and breaking down the problem into smaller problems, difficulty finding the features the programs offer, or documentation problems.

Yeh and Kim~\cite{yeh_craftml_2018} report problems with programming-based CAD and direct manipulation software offering scripting features for 3D design.
These problems include difficulties reading code, re-using code, aligning objects, selecting parts, refactoring code, and 3D printing.
Unfortunately, these undetailed findings were obtained from undocumented feedback from novice students with OpenSCAD and online forums such as StackOverflow.
More recently, Gonzalez \textit{et al.}~\cite{gonzalez_introducing_2023} investigated challenges in OpenSCAD related to navigating and editing models.
The findings include difficulties in linking the code to the view and performing spatial transformations.
We aim to cover a broader scope of the 3D design, 3D printing, and re-using pre-existing models experience with OpenSCAD.

\subsection{Sharing and re-using models.}

Sharing is a keystone in the 3D printing community \cite{kuznetsov_rise_2010}.
Multiple model-storing websites allow authors to upload models to share with other users, such as Thingiverse \cite{thingiversecom_thingiverse_2022}, MyMiniFactory \cite{myminifactory_myminifactory_2022} or Printables \cite{prusa_research_as_base_2023}.
Some of them allow authors to upload coded parametric models that expose widgets so other users can modify parameter values for the system to create a customized version of the models \cite{shapiro_makewithtech_2023,thingiversecom_customizer_2022}.
Thingiverse, with its Customizer application \cite{thingiversecom_customizer_2022} is an example of these solutions that store OpenSCAD models.
Oehlberg \textit{et al.}~\cite{oehlberg_patterns_2015} reported how, after a year of the release of Customizer, about 40\% of the Thingiverse models were created from parametric models.
These findings depict how some users re-use models to remix them by changing parameters.
However, it is unknown if the authors of these parametric models adopt the re-using practice in their models.
We aim to understand the role of model-storing websites in programming-based CAD design.

In summary, some previous research has explored end-user experiences in direct manipulation programs, with findings not always applicable to programming-based CAD.
Furthermore, there has been limited exploration of programming-based CAD problems and the sharing and re-using practice.
Our work addresses these voids and contributes to a better understanding of this population in depth.

\section{Method} \label{method}

We conducted twenty semi-structured interviews to understand the motivations and challenges of OpenSCAD users empirically. The interview was divided into three parts.
First, we asked participants for demographic information.
Also, we asked them to self-rate their skill level on a scale from one to five, one meaning novice and five expert, on direct manipulation CAD applications, programming-based CAD applications, and general programming languages outside CAD.
Similarly, we asked participants to self-rate their skill level in OpenSCAD on the same scale.
The responses are reported in Table~\ref{tab:demographics}.

In the second part, we asked participants open questions about their experience in 3D printing and 3D modeling.
Specifically, we were interested in understanding the motivations of makers in using OpenSCAD for 3D design, the challenges and limitations they face using OpenSCAD for 3D printing, their perception of direct manipulation programs compared to OpenSCAD, and ideas to improve OpenSCAD that might apply to other programming-based CAD applications.
Furthermore, we draw on previous work on understanding the complexity and challenges of 3D modeling in direct manipulation programs to contrast these findings with the experience of OpenSCAD users. Concretely, we have included questions related to difficulties measuring physical objects and transferring data to digital designs \cite{mahapatra_barriers_2019}, and sharing and re-using models in model-storing websites \cite{alcock_barriers_2016,yeh_craftml_2018}.

Finally, we wanted to understand the limitations of more specific actions when designing in OpenSCAD.
We decided it would be easier to study participants' behavior in a real scenario while we observe them instead of only asking them to describe how they use the software, which could lead to easily missing specific actions or strategies they use.
Thus, we asked the participants to perform a short hands-on exercise in the third part to observe their behavior while performing tasks.
Based on the findings of previous work, we report problems in programming-based CAD related to selecting specific parts to apply operations, including challenges in reading, navigating, refactoring, and understanding code~\cite{yeh_craftml_2018,gonzalez_introducing_2023}.
If possible, we asked the participants to bring one of their own OpenSCAD models to the interview.
P2 did not provide a model, so we used the example \texttt{candleStand.scad} provided by OpenSCAD.

The participants explained the motivation behind the model and went through their code, discussing difficulties and how they modeled their object.
We asked them to perform search tasks replicating the need to select a part to modify it or apply an operation.
We pointed at specific parts in the 3D view and asked the participant to locate the lines of code that created them.
We asked participants to think aloud while we carefully observed the process, recurrent behaviors, and strategies.
We paid special attention to the software features they used, the typical patterns they followed to perform the tasks, and the errors they made.
Last, we discussed ideas they could have to improve their experiences in OpenSCAD. 

The interviews lasted approximately 60 minutes on average.
The questionnaire used is included in the Appendix. We took notes of their answers and the observed behaviors during the hands-on exercise.
The experiment protocol was examined and approved by the ethics board in our laboratory.

\aptLtoX[graphic=no,type=html]{\begin{table}
\Description{Demographic information of the participants. The participant's id is placed in the first column, starting with P1 and ending with P2. The second column stores the age range of participants split into 20 to 29, 30 to 39, 40 to 49, 50 to 59, and 60 to 69. The third column stores the 3D printing experience of the participants in years. The fourth column stores the self-rated skill level in OpenSCAD. The fifth to the tenth columns store the participants' self-rated skill level in the direct manipulation tools Blender, FreeCAD, Fusion 360, TinkerCAD, Rhinoceros, and Others. The eleventh to the sixteenth columns store participants' self-rated skill level in the programming-based CAD tools IceSL, Python API, JsCAD, CadQuery, BRLCAD, and BlocksCAD. The twelfth to the nineteenth columns store participants' self-rated skill levels in the programming languages of general-purpose C++, C\#, Java, JavaScript, Python, PHP, and Others}
\caption[Caption for demogrphics]{Demographics and self-rated skill level in CAD programs and programming languages. \\ Participants self-rated their skill level on the scale: 1 (Novice), 2 (Advanced Beginner), 3 (Competent), 4 (Proficient), 5 (Expert). The level reported in the category \emph{Others} is the highest rank expressed by the participant among the options.
\\ *Direct manipulation CAD others: LibreCAD, Sketch Up, AutoCAD, Curve3D, OnShape, Catia, SolidWorks.
\\ **Programming Language Others: Prolog, MaxMSP, PureData, Ruby, GoLink, MatLab, Cobol, Pearl, Pascal, Groovy, TypeScript.}
\label{tab:demographics}
\footnotesize
\centering

\begin{tabular}{l!{\color{black}\vrule}clc!{\color{black}\vrule}llllll!{\color{black}\vrule}llllll!{\color{black}\vrule}llllllll!{\color{black}\vrule}}
\arrayrulecolor{black}\cline{5-24}
\multicolumn{1}{l}{} &              &                        & \multicolumn{1}{l!{\color{black}\vrule}}{}              & \multicolumn{6}{c!{\color{black}\vrule}}{Direct Manipulation CAD}                      & \multicolumn{6}{c!{\color{black}\vrule}}{Programming based CAD}            & \multicolumn{8}{c!{\color{black}\vrule}}{Programming languages}    \\
\arrayrulecolor{black}\hline
\multicolumn{1}{|l|}{\rotatebox[origin=c]{90}{Participant}} & \rotatebox[origin=c]{90}{Age Range}& \rotatebox[origin=c]{90}{ 3D printing } \rotatebox[origin=c]{90}{ experience (y)  }& \rotatebox[origin=c]{90}{OpenSCAD} & \rotatebox[origin=c]{90}{Blender}& \rotatebox[origin=c]{90}{FreeCAD}& \rotatebox[origin=c]{90}{Fusion 360}& \rotatebox[origin=c]{90}{TinkerCAD}& \rotatebox[origin=c]{90}{Rhinoceros}& \rotatebox[origin=c]{90}{Others*}& \rotatebox[origin=c]{90}{IceSL}& \rotatebox[origin=c]{90}{Python API}&  \rotatebox[origin=c]{90}{JsCAD}& \rotatebox[origin=c]{90}{CadQuery}& \rotatebox[origin=c]{90}{BRLCAD}& \rotatebox[origin=c]{90}{BlocksCAD}&  \rotatebox[origin=c]{90}{C}& \rotatebox[origin=c]{90}{C++}& \rotatebox[origin=c]{90}{C\#}& \rotatebox[origin=c]{90}{Java}& \rotatebox[origin=c]{90}{JavaScript}& \rotatebox[origin=c]{90}{Python}& \rotatebox[origin=c]{90}{PHP}& \rotatebox[origin=c]{90}{Others**}\\
\arrayrulecolor{black}\hline
\arrayrulecolor{black}\multicolumn{1}{|l|}{P1}                   & 50 - 59      & 10                     &  \cellcolor[HTML]{B6DCA6}{3}                                                      & \cellcolor[HTML]{F1F8ED}{1}        & \cellcolor[HTML]{F1F8ED}{1}        &           &            &         & \cellcolor[HTML]{F1F8ED}{1}       &            &               &       &          &        &           & \cellcolor[HTML]{4DA747}{4} &     &     & \cellcolor[HTML]{4DA747}{4}    & \cellcolor[HTML]{4DA747}{4}     & \cellcolor[HTML]{4DA747}{4}      & \cellcolor[HTML]{4DA747}{4}   &  \cellcolor[HTML]{4DA747}{4}        \\
\hline
\arrayrulecolor{black}\multicolumn{1}{|l|}{P2}                   & 40 - 49      & 6                      & \cellcolor[HTML]{F1F8ED}{1}                                                       & \cellcolor[HTML]{B6DCA6}{3}       &            &           &            &         &        & \cellcolor[HTML]{598F55}{5}     & \cellcolor[HTML]{E9FaE0}{2}          &       &          &        &           &   & \cellcolor[HTML]{598F55}{5}   &     &      & \cellcolor[HTML]{B6DCA6}{3}          & \cellcolor[HTML]{4DA747}{4}      &     & \cellcolor[HTML]{B6DCA6}{3}       \\
\hline
\arrayrulecolor{black}\multicolumn{1}{|l|}{P3}                   & 20 - 29      & 7                      &  \cellcolor[HTML]{4DA747}{4}                                                       &         & \cellcolor[HTML]{E9FAE0}{2}       & \cellcolor[HTML]{4DA747}{4}          &            &         & \cellcolor[HTML]{F1F8ED}{1}       &            &               &       &          &        &           &   & \cellcolor[HTML]{4DA747}{4}   &     &      &            & \cellcolor[HTML]{4DA747}{4}      &     & \cellcolor[HTML]{598F55}{5}       \\
\hline
\arrayrulecolor{black}\multicolumn{1}{|l|}{P4}                   & 30 - 39      &  5                      &  \cellcolor[HTML]{B6DCA6}{3}                                                       &         & \cellcolor[HTML]{B6DCA6}{3}       & \cellcolor[HTML]{B6DCA6}{3}          &            &         &        &       &            &       &          &        &           & \cellcolor[HTML]{4DA747}{4} & \cellcolor[HTML]{4DA747}{4}   & \cellcolor[HTML]{4DA747}{4}   & \cellcolor[HTML]{4DA747}{4}    &            & \cellcolor[HTML]{4DA747}{4}      &     & \cellcolor[HTML]{4DA747}{4}       \\
\hline
\arrayrulecolor{black}\multicolumn{1}{|l|}{P5}                   & 40 - 49     & 6                      &  \cellcolor[HTML]{4DA747}{4}                                                       &         &         &            &            &         & \cellcolor[HTML]{4DA747}{4}         &       &               &       &          &        &           & \cellcolor[HTML]{4DA747}{4} &     & \cellcolor[HTML]{4DA747}{4}   & \cellcolor[HTML]{4DA747}{4}    &            &        &        & \cellcolor[HTML]{B6DCA6}{3}       \\
\hline
\arrayrulecolor{black}\multicolumn{1}{|l|}{P6}                   & 40 - 49      & 15                     &  \cellcolor[HTML]{4DA747}{4}                                                       &         &         &            & \cellcolor[HTML]{E9FaE0}{2}         &         &        &           &               &       &          &        &           &   &     & \cellcolor[HTML]{4DA747}{4}   & \cellcolor[HTML]{E9FaE0}{2}    &            & \cellcolor[HTML]{4DA747}{4}      & \cellcolor[HTML]{B6DCA6}{3}   &         \\
\hline
\arrayrulecolor{black}\multicolumn{1}{|l|}{P7}                   & 30 - 39      & 8                      &  \cellcolor[HTML]{4DA747}{4}                                                       &         &         & \cellcolor[HTML]{B6DCA6}{3}          &            &         &        &       &               & \cellcolor[HTML]{B6DCA6}{3}     & \cellcolor[HTML]{B6DCA6}{3}        &        &           & \cellcolor[HTML]{4DA747}{4} &     &     & \cellcolor[HTML]{4DA747}{4}    &            & \cellcolor[HTML]{4DA747}{4}      &     &         \\
\hline
\arrayrulecolor{black}\multicolumn{1}{|l|}{P8}                   & 40 - 49      & 1.5                    &  \cellcolor[HTML]{B6DCA6}{3}                                                       &         &         &            & \cellcolor[HTML]{E9FaE0}{2}         &         &        &            &               &       &          &        &           & \cellcolor[HTML]{4DA747}{4} & \cellcolor[HTML]{4DA747}{4}   &     &      & \cellcolor[HTML]{598F55}{5}          &        &     & \cellcolor[HTML]{598F55}{5}       \\
\hline
\arrayrulecolor{black}\multicolumn{1}{|l|}{P9}                   & 30 - 39      & 14                     &  \cellcolor[HTML]{B6DCA6}{3}                                                       & \cellcolor[HTML]{4DA747}{4}       & \cellcolor[HTML]{B6DCA6}{3}       &            &           & \cellcolor[HTML]{4DA747}{4}          & \cellcolor[HTML]{B6DCA6}{3}      &            &               &       &          &        &           &   & \cellcolor[HTML]{4DA747}{4}   &     &      & \cellcolor[HTML]{598F55}{5}          & \cellcolor[HTML]{B6DCA6}{3}      &     & \cellcolor[HTML]{B6DCA6}{3}       \\
\hline
\arrayrulecolor{black}\multicolumn{1}{|l|}{P10}                  & 60 - 69      & 8                      & \cellcolor[HTML]{598F55}{5}                                                       &         &         &            & \cellcolor[HTML]{E9FaE0}{2}         & \cellcolor[HTML]{F1F8ED}{1}           &        &       &               &       &          &        &           &   & \cellcolor[HTML]{598F55}{5}   &     &      &            & \cellcolor[HTML]{598F55}{5}      &     & \cellcolor[HTML]{598F55}{5}       \\
\hline
\arrayrulecolor{black}\multicolumn{1}{|l|}{P11}                  & 40 - 49      & 7                      & \cellcolor[HTML]{4DA747}{4}                                                       &         & \cellcolor[HTML]{F1F8ED}{1}        &           &            &         &        &            &               & \cellcolor[HTML]{F1F8ED}{1}      &          &        &           &   &     &     &      & \cellcolor[HTML]{598F55}{5}          & \cellcolor[HTML]{598F55}{5}      & \cellcolor[HTML]{598F55}{5}   &         \\
\hline
\arrayrulecolor{black}\multicolumn{1}{|l|}{P12}                  & 50 - 59      & 13                     &  \cellcolor[HTML]{4DA747}{4}                                                       &         & \cellcolor[HTML]{F1F8ED}{1}        &           & \cellcolor[HTML]{F1F8ED}{1}          &            & \cellcolor[HTML]{F1F8ED}{1}       &       &               & \cellcolor[HTML]{4DA747}{4}     &          & \cellcolor[HTML]{4DA747}{4}      & \cellcolor[HTML]{4DA747}{4}         &   &     &     &      &            &        &     & \cellcolor[HTML]{4DA747}{4}       \\
\hline
\arrayrulecolor{black}\multicolumn{1}{|l|}{P13}                  & 60 - 69     &  5                      &  \cellcolor[HTML]{B6DCA6}{3}                                                       &         & \cellcolor[HTML]{F1F8ED}{1}        & \cellcolor[HTML]{E9FaE0}{2}          &            &         &        &       &            &       & \cellcolor[HTML]{F1F8ED}{1}         &        &           & \cellcolor[HTML]{F1F8ED}{1}  &     &     &      &        & \cellcolor[HTML]{F1F8ED}{1}       &     &         \\
\hline
\arrayrulecolor{black}\multicolumn{1}{|l|}{P14}                  & 60 - 69      &  4                      &  \cellcolor[HTML]{598F55}{5}                                                       & \cellcolor[HTML]{F1F8ED}{1}        & \cellcolor[HTML]{B6DCA6}{3}       &            & \cellcolor[HTML]{598F55}{5}         &         &       &            &               &       &          &        &           &   &     &     & \cellcolor[HTML]{E9FaE0}{2}    & \cellcolor[HTML]{4DA747}{4}          &        & \cellcolor[HTML]{4DA747}{4}   & \cellcolor[HTML]{B6DCA6}{3}       \\
\hline
\arrayrulecolor{black}\multicolumn{1}{|l|}{P15}                  & 40 - 49      & 8                      &  \cellcolor[HTML]{B6DCA6}{3}                                                       & \cellcolor[HTML]{F1F8ED}{1}        &         &            & \cellcolor[HTML]{F1F8ED}{1}          &         & \cellcolor[HTML]{F1F8ED}{1}           &       &               &       &          &        &           &   &     &     &      & \cellcolor[HTML]{4DA747}{4}          &\cellcolor[HTML]{4DA747}{4}      &     & \cellcolor[HTML]{4DA747}{4}       \\
\hline
\arrayrulecolor{black}\multicolumn{1}{|l|}{P16}                  & 50 - 59      & 6                      &  \cellcolor[HTML]{4DA747}{4}                                                       &         &         &            &           &            & \cellcolor[HTML]{F1F8ED}{1}        &       &            &               &          &        &           & \cellcolor[HTML]{4DA747}{4} & \cellcolor[HTML]{4DA747}{4}   &     &      & \cellcolor[HTML]{4DA747}{4}          & \cellcolor[HTML]{4DA747}{4}      &     &         \\
\hline
\arrayrulecolor{black}\multicolumn{1}{|l|}{P17}                  & 30 - 39      & 15                     &  \cellcolor[HTML]{598F55}{5}                                                       & \cellcolor[HTML]{4DA747}{4}       &         & \cellcolor[HTML]{4DA747}{4}          & \cellcolor[HTML]{4DA747}{4}         &         &        &       &            &       &          &        &           & \cellcolor[HTML]{E9FaE0}{2} & \cellcolor[HTML]{E9FaE0}{2}   &     &      & \cellcolor[HTML]{4DA747}{4}          & \cellcolor[HTML]{4DA747}{4}      &     &         \\
\hline
\arrayrulecolor{black}\multicolumn{1}{|l|}{P18}                  & 60 - 69      & 8                      &  \cellcolor[HTML]{E9FaE0}{2}                                                       &         & \cellcolor[HTML]{F1F8ED}{1}        & \cellcolor[HTML]{F1F8ED}{1}           & \cellcolor[HTML]{F1F8ED}{1}          &         &      &            &               &       &          &        &           &   &     &     &      &            &        &     &         \\
\hline
\arrayrulecolor{black}\multicolumn{1}{|l|}{P19}                  & 40 - 49      &  3                      &  \cellcolor[HTML]{4DA747}{4}                                                       &         &         & \cellcolor[HTML]{F1F8ED}{1}           & \cellcolor[HTML]{F1F8ED}{1}          &            &        &           &               &       &          &        &           &   &     &     & \cellcolor[HTML]{598F55}{5}    &            & \cellcolor[HTML]{598F55}{5}      &     & \cellcolor[HTML]{B6DCA6}{3}       \\
\hline
\arrayrulecolor{black}\multicolumn{1}{|l|}{P20}                  & 50 - 59      & 9                      & {\cellcolor{skillRate5}} 5                                                       &         &            &           &           &         &        &       & \cellcolor[HTML]{F1F8ED}{1}           &       &          &        &           &   &     & \cellcolor[HTML]{4DA747}{4}   &      &            & \cellcolor[HTML]{4DA747}{4}      &     & \cellcolor[HTML]{598F55}{5}       \\
\arrayrulecolor{black}\hline
\end{tabular}
\end{table}}{\begin{table*}
\Description{Demographic information of the participants. The participant's id is placed in the first column, starting with P1 and ending with P2. The second column stores the age range of participants split into 20 to 29, 30 to 39, 40 to 49, 50 to 59, and 60 to 69. The third column stores the 3D printing experience of the participants in years. The fourth column stores the self-rated skill level in OpenSCAD. The fifth to the tenth columns store the participants' self-rated skill level in the direct manipulation tools Blender, FreeCAD, Fusion 360, TinkerCAD, Rhinoceros, and Others. The eleventh to the sixteenth columns store participants' self-rated skill level in the programming-based CAD tools IceSL, Python API, JsCAD, CadQuery, BRLCAD, and BlocksCAD. The twelfth to the nineteenth columns store participants' self-rated skill levels in the programming languages of general-purpose C++, C\#, Java, JavaScript, Python, PHP, and Others}
\caption[Caption for demogrphics]{Demographics and self-rated skill level in CAD programs and programming languages. \\ Participants self-rated their skill level on the scale: 1 (Novice), 2 (Advanced Beginner), 3 (Competent), 4 (Proficient), 5 (Expert). The level reported in the category \emph{Others} is the highest rank expressed by the participant among the options.
\\ *Direct manipulation CAD others: LibreCAD, Sketch Up, AutoCAD, Curve3D, OnShape, Catia, SolidWorks.
\\ **Programming Language Others: Prolog, MaxMSP, PureData, Ruby, GoLink, MatLab, Cobol, Pearl, Pascal, Groovy, TypeScript.}
\label{tab:demographics}
\footnotesize
\centering
\begin{tabular}{l!{\color{black}\vrule}clc!{\color{black}\vrule}llllll!{\color{black}\vrule}llllll!{\color{black}\vrule}llllllll!{\color{black}\vrule}}
\arrayrulecolor{black}\cline{5-24}
\multicolumn{1}{l}{} &              &                        & \multicolumn{1}{l!{\color{black}\vrule}}{}              & \multicolumn{6}{c!{\color{black}\vrule}}{Direct Manipulation CAD}                      & \multicolumn{6}{c!{\color{black}\vrule}}{Programming based CAD}            & \multicolumn{8}{c!{\color{black}\vrule}}{Programming languages}    \\
\arrayrulecolor{black}\hline
\multicolumn{1}{|l|}{\rotatebox[origin=c]{90}{Participant}} & \rotatebox[origin=c]{90}{Age Range}& \rotatebox[origin=c]{90}{ 3D printing } \rotatebox[origin=c]{90}{ experience (y)  }& \rotatebox[origin=c]{90}{OpenSCAD} & \rotatebox[origin=c]{90}{Blender}& \rotatebox[origin=c]{90}{FreeCAD}& \rotatebox[origin=c]{90}{Fusion 360}& \rotatebox[origin=c]{90}{TinkerCAD}& \rotatebox[origin=c]{90}{Rhinoceros}& \rotatebox[origin=c]{90}{Others*}& \rotatebox[origin=c]{90}{IceSL}& \rotatebox[origin=c]{90}{Python API}&  \rotatebox[origin=c]{90}{JsCAD}& \rotatebox[origin=c]{90}{CadQuery}& \rotatebox[origin=c]{90}{BRLCAD}& \rotatebox[origin=c]{90}{BlocksCAD}&  \rotatebox[origin=c]{90}{C}& \rotatebox[origin=c]{90}{C++}& \rotatebox[origin=c]{90}{C\#}& \rotatebox[origin=c]{90}{Java}& \rotatebox[origin=c]{90}{JavaScript}& \rotatebox[origin=c]{90}{Python}& \rotatebox[origin=c]{90}{PHP}& \rotatebox[origin=c]{90}{Others**}\\
\arrayrulecolor{black}\hline
\arrayrulecolor{black}\multicolumn{1}{|l|}{P1}                   & 50 - 59      & 10                     &  {\cellcolor{skillRate3}}3                                                       & {\cellcolor{skillRate1}}1        & {\cellcolor{skillRate1}}1        &           &            &         & {\cellcolor{skillRate1}}1       &            &               &       &          &        &           & {\cellcolor{skillRate4}}4 &     &     & {\cellcolor{skillRate4}}4    & {\cellcolor{skillRate4}}4     & {\cellcolor{skillRate4}}4      & {\cellcolor{skillRate4}}4   &  {\cellcolor{skillRate4}}4        \\
\hline
\arrayrulecolor{black}\multicolumn{1}{|l|}{P2}                   & 40 - 49      & 6                      & {\cellcolor{skillRate1}}1                                                       & {\cellcolor{skillRate3}}3       &            &           &            &         &        & {\cellcolor{skillRate5}}5     & {\cellcolor{skillRate2}}2          &       &          &        &           &   & {\cellcolor{skillRate5}}5   &     &      & {\cellcolor{skillRate3}}3          & {\cellcolor{skillRate4}}4      &     & {\cellcolor{skillRate3}}3       \\
\hline
\arrayrulecolor{black}\multicolumn{1}{|l|}{P3}                   & 20 - 29      & 7                      &  {\cellcolor{skillRate4}}4                                                       &         & {\cellcolor{skillRate2}}2       & {\cellcolor{skillRate4}}4          &            &         & {\cellcolor{skillRate1}}1       &            &               &       &          &        &           &   & {\cellcolor{skillRate4}}4   &     &      &            & {\cellcolor{skillRate4}}4      &     & {\cellcolor{skillRate5}}5       \\
\hline
\arrayrulecolor{black}\multicolumn{1}{|l|}{P4}                   & 30 - 39      &  5                      &  {\cellcolor{skillRate3}}3                                                       &         & {\cellcolor{skillRate3}}3       & {\cellcolor{skillRate3}}3          &            &         &        &       &            &       &          &        &           & {\cellcolor{skillRate4}}4 & {\cellcolor{skillRate4}}4   & {\cellcolor{skillRate4}}4   & {\cellcolor{skillRate4}}4    &            & {\cellcolor{skillRate4}}4      &     & {\cellcolor{skillRate4}}4       \\
\hline
\arrayrulecolor{black}\multicolumn{1}{|l|}{P5}                   & 40 - 49     & 6                      &  {\cellcolor{skillRate4}}4                                                       &         &         &            &            &         & {\cellcolor{skillRate4}}4         &       &               &       &          &        &           & {\cellcolor{skillRate4}}4 &     & {\cellcolor{skillRate4}}4   & {\cellcolor{skillRate4}}4    &            &        &        & {\cellcolor{skillRate3}}3       \\
\hline
\arrayrulecolor{black}\multicolumn{1}{|l|}{P6}                   & 40 - 49      & 15                     &  {\cellcolor{skillRate4}}4                                                       &         &         &            & {\cellcolor{skillRate2}}2         &         &        &           &               &       &          &        &           &   &     & {\cellcolor{skillRate4}}4   & {\cellcolor{skillRate2}}2    &            & {\cellcolor{skillRate4}}4      & {\cellcolor{skillRate3}}3   &         \\
\hline
\arrayrulecolor{black}\multicolumn{1}{|l|}{P7}                   & 30 - 39      & 8                      &  {\cellcolor{skillRate4}}4                                                       &         &         & {\cellcolor{skillRate3}}3          &            &         &        &       &               & {\cellcolor{skillRate3}}3     & {\cellcolor{skillRate3}}3        &        &           & {\cellcolor{skillRate4}}4 &     &     & {\cellcolor{skillRate4}}4    &            & {\cellcolor{skillRate4}}4      &     &         \\
\hline
\arrayrulecolor{black}\multicolumn{1}{|l|}{P8}                   & 40 - 49      & 1.5                    &  {\cellcolor{skillRate3}}3                                                       &         &         &            & {\cellcolor{skillRate2}}2         &         &        &            &               &       &          &        &           & {\cellcolor{skillRate4}}4 & {\cellcolor{skillRate4}}4   &     &      & {\cellcolor{skillRate5}}5          &        &     & {\cellcolor{skillRate5}}5       \\
\hline
\arrayrulecolor{black}\multicolumn{1}{|l|}{P9}                   & 30 - 39      & 14                     &  {\cellcolor{skillRate3}}3                                                       & {\cellcolor{skillRate4}}4       & {\cellcolor{skillRate3}}3       &            &           & {\cellcolor{skillRate4}}4          & {\cellcolor{skillRate3}}3      &            &               &       &          &        &           &   & {\cellcolor{skillRate4}}4   &     &      & {\cellcolor{skillRate5}}5          & {\cellcolor{skillRate3}}3      &     & {\cellcolor{skillRate3}}3       \\
\hline
\arrayrulecolor{black}\multicolumn{1}{|l|}{P10}                  & 60 - 69      & 8                      & {\cellcolor{skillRate5}}5                                                       &         &         &            & {\cellcolor{skillRate2}}2         & {\cellcolor{skillRate1}}1           &        &       &               &       &          &        &           &   & {\cellcolor{skillRate5}}5   &     &      &            & {\cellcolor{skillRate5}}5      &     & {\cellcolor{skillRate5}}5       \\
\hline
\arrayrulecolor{black}\multicolumn{1}{|l|}{P11}                  & 40 - 49      & 7                      & {\cellcolor{skillRate4}}4                                                       &         & {\cellcolor{skillRate1}}1        &           &            &         &        &            &               & {\cellcolor{skillRate1}}1      &          &        &           &   &     &     &      & {\cellcolor{skillRate5}}5          & {\cellcolor{skillRate5}}5      & {\cellcolor{skillRate5}}5   &         \\
\hline
\arrayrulecolor{black}\multicolumn{1}{|l|}{P12}                  & 50 - 59      & 13                     &  {\cellcolor{skillRate4}}4                                                       &         & {\cellcolor{skillRate1}}1        &           & {\cellcolor{skillRate1}}1          &            & {\cellcolor{skillRate1}}1       &       &               & {\cellcolor{skillRate4}}4     &          & {\cellcolor{skillRate4}}4      & {\cellcolor{skillRate4}}4         &   &     &     &      &            &        &     & {\cellcolor{skillRate4}}4       \\
\hline
\arrayrulecolor{black}\multicolumn{1}{|l|}{P13}                  & 60 - 69     &  5                      &  {\cellcolor{skillRate3}}3                                                       &         & {\cellcolor{skillRate1}}1        & {\cellcolor{skillRate2}}2          &            &         &        &       &            &       & {\cellcolor{skillRate1}}1         &        &           & {\cellcolor{skillRate1}}1  &     &     &      &        & {\cellcolor{skillRate1}}1       &     &         \\
\hline
\arrayrulecolor{black}\multicolumn{1}{|l|}{P14}                  & 60 - 69      &  4                      &  {\cellcolor{skillRate5}}5                                                       & {\cellcolor{skillRate1}}1        & {\cellcolor{skillRate3}}3       &            & {\cellcolor{skillRate5}}5         &         &       &            &               &       &          &        &           &   &     &     & {\cellcolor{skillRate2}}2    & {\cellcolor{skillRate4}}4          &        & {\cellcolor{skillRate4}}4   & {\cellcolor{skillRate3}}3       \\
\hline
\arrayrulecolor{black}\multicolumn{1}{|l|}{P15}                  & 40 - 49      & 8                      &  {\cellcolor{skillRate3}}3                                                       & {\cellcolor{skillRate1}}1        &         &            & {\cellcolor{skillRate1}}1          &         & {\cellcolor{skillRate1}}1           &       &               &       &          &        &           &   &     &     &      & {\cellcolor{skillRate4}}4          &{\cellcolor{skillRate4}}4      &     & {\cellcolor{skillRate4}}4       \\
\hline
\arrayrulecolor{black}\multicolumn{1}{|l|}{P16}                  & 50 - 59      & 6                      &  {\cellcolor{skillRate4}}4                                                       &         &         &            &           &            & {\cellcolor{skillRate1}}1        &       &            &               &          &        &           & {\cellcolor{skillRate4}}4 & {\cellcolor{skillRate4}}4   &     &      & {\cellcolor{skillRate4}}4          & {\cellcolor{skillRate4}}4      &     &         \\
\hline
\arrayrulecolor{black}\multicolumn{1}{|l|}{P17}                  & 30 - 39      & 15                     &  {\cellcolor{skillRate5}}5                                                       & {\cellcolor{skillRate4}}4       &         & {\cellcolor{skillRate4}}4          & {\cellcolor{skillRate4}}4         &         &        &       &            &       &          &        &           & {\cellcolor{skillRate2}}2 & {\cellcolor{skillRate2}}2   &     &      & {\cellcolor{skillRate4}}4          & {\cellcolor{skillRate4}}4      &     &         \\
\hline
\arrayrulecolor{black}\multicolumn{1}{|l|}{P18}                  & 60 - 69      & 8                      &  {\cellcolor{skillRate2}}2                                                       &         & {\cellcolor{skillRate1}}1        & {\cellcolor{skillRate1}}1           & {\cellcolor{skillRate1}}1          &         &      &            &               &       &          &        &           &   &     &     &      &            &        &     &         \\
\hline
\arrayrulecolor{black}\multicolumn{1}{|l|}{P19}                  & 40 - 49      &  3                      &  {\cellcolor{skillRate4}}4                                                       &         &         & {\cellcolor{skillRate1}}1           & {\cellcolor{skillRate1}}1          &            &        &           &               &       &          &        &           &   &     &     & {\cellcolor{skillRate5}}5    &            & {\cellcolor{skillRate5}}5      &     & {\cellcolor{skillRate3}}3       \\
\hline
\arrayrulecolor{black}\multicolumn{1}{|l|}{P20}                  & 50 - 59      & 9                      & {\cellcolor{skillRate5}} 5                                                       &         &            &           &           &         &        &       & {\cellcolor{skillRate1}}1           &       &          &        &           &   &     & {\cellcolor{skillRate4}}4   &      &            & {\cellcolor{skillRate4}}4      &     & {\cellcolor{skillRate5}}5       \\
\arrayrulecolor{black}\hline
\end{tabular}
\end{table*}}

\subsection{Recruitment and Participants}

We relied on the common use of 3D modeling in research and the active sharing nature of programming and maker communities on social media.
We recruited participants from research laboratories and OpenSCAD channels on Reddit (\verb+r/openscad+) and Facebook (\verb+OpenSCADAcademy+) to conduct the semi-structured interview using video conferencing or in person.
The only requirement was having enough experience with OpenSCAD to read and write code, but we also expressed that having 3D printing experience would be an asset.

We report participant demographics and experience in Table~\ref{tab:demographics}.
All the participants self-identified as male and varied in age: one was between 20 and 29, four were between 30 and 39, seven were between 40 and 49, four were between 50 and 59, and four were between 60 and 69 (average: 48.0, standard deviation: 11.7).
All participants, except P8, had three or more years of 3D printing experience (average: 7.9y, standard deviation: 3.8).
Except for P13 and P18, all participants self-rated with four or more in at least two programming languages.
Moreover, all participants, except P20 mentioned having experience with direct manipulation CAD programs. Only five participants self-rated their direct manipulation CAD application skills with four or more.
Regarding experience with other programming-based CAD applications or applications that allow scripting, only six participants expressed having any, and only P2 and P12 self-rated their skill level above 3 in one of those applications.
Finally, participants self-rated their skill level with OpenSCAD as follows: One participant with 1, one participant with 2, six participants with 3, eight participants with 4, and four participants with 5.

\subsection{Data Analysis}

We followed a Reflexive Thematic Analysis (RTA)~\cite{braun_successful_2013,byrne_worked_2022} approach in an iterative coding process.
Our study aims to understand the user experience using OpenSCAD, and part of the data collected included behavioral observations from the hands-on experience.
Thus, we opted for a data analysis approach suitable for these studies~\cite{braun_can_2021,byrne_worked_2022} that allows flexible participation of the researcher's interpretations rather than other qualitative analysis approaches such as code reliability or ground theory \cite{boyatzis_transforming_1998,smith_qualitative_2011}.

We uploaded the interview data into the MaxQDA data analysis software~\cite{verbi_gmbh_software_2023}.
One of the researchers performed an inductive analysis to develop a set of codes by coding the first ten interviews.
Then, the coder started grouping codes by recognizing recurring patterns and identifying codes describing a central concept to create subthemes and themes~\cite{braun_answers_2019}.
To achieve a richer interpretation of the coding process~\cite{braun_can_2021}, a second researcher performed a deductive thematic analysis on a randomly selected interview.
The second coder used the codes created by the first coder in this interview and could create new codes when necessary.
Then, both coders discussed the disagreements and refined the codes by removing, merging, changing, or adding new codes.
After re-organizing codes, subthemes, and themes, the first ten interviews were re-coded with the resulting set of codes.
In the second iteration, the first coder continued the inductive analysis with the next five interviews, followed by the deductive coding from the second coder, discussions on the codes, refinement of the codebook, and re-coding of the interviews.
A third iteration was performed to complete the coding of the total of interviews.

Although RTA does not seek reliability coding~\cite{braun_successful_2013}, we were interested in tracking the level of agreement between coders.
We calculated Cohen's kappa index in every iteration to verify inter-coder reliability~\cite{mcdonald_reliability_2019}.
At the end of the coding of all interviews, we achieved a \textit{substantial}~\cite{landis_measurement_1977} agreement: iteration 1 \(\kappa = 0.543\), iteration 2 \(\kappa = 0.592\), iteration 3 \textbf{\(\kappa = 0.617\)}.

Most of the codes were created in the first fourteen interviews, achieving a potential code saturation. However, it was not until the seventeenth interview that codes, themes, and subthemes found in the codebook did not have substantial changes, and their meaning was well established, achieving meaning saturation \cite{hennink_code_2017}.

\section{Themes}

We created 266 individual codes to code a total of 783 segments of our notes.
We grouped codes into subthemes and then into three main themes.
\textit{Programming-based CAD user profile} theme groups 22 codes (59 segments coded) and 9 subthemes as depicted in Table \ref{tab:Theme1}.
Table \ref{tab:Theme2} depicts the theme \textit{Design} that covers 193 codes (632 segments coded) and 80 subthemes.
Finally, \textit{Printing} theme includes 51 codes (92 segments coded) and 20 subthemes (Table \ref{tab:Theme3}).

\subsection{Programming-based CAD users profile} \label{sec::theme1_profile}

We start by discussing the design experience of the participants in OpenSCAD.
Later, we discuss why participants use a programming-based CAD and their opinions on direct manipulation programs.

\begin{table*}[ht]
\Description{The table depicts the structure of subthemes of the theme Programming-based CAD user profile. From the left to the right, the table stores subthemes with their names and the number of participants who commented on them. In the left column are the most specific subthemes. Moving to the right, subthemes are grouped in mode general subthemes until reaching the theme in the most right part of the table. Themes are green colored with different intensities proportional to the number of participants that commented on the subtheme or theme.}
\caption{Structure of theme \textit{Programming-based CAD user profile}. Color intensity is proportional to the number of interviews coded with codes of the theme and subthemes.}
\label{tab:Theme1}
\centering
\includegraphics[trim={1.8cm 22cm 2.2cm 1.8cm}, clip, width=.98\textwidth]{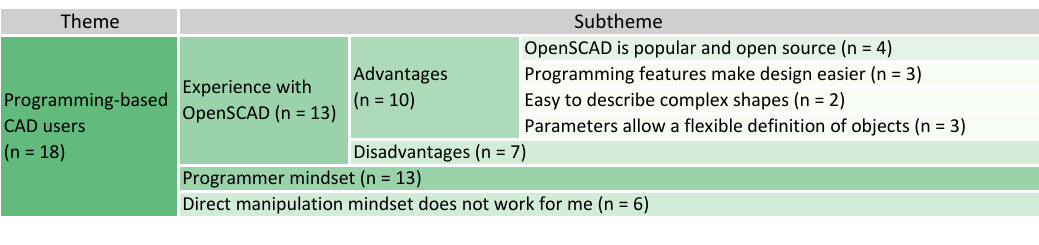}
\end{table*}

\subsubsection{Experience with OpenSCAD}

Participants (\textit{n = 10}) mentioned the advantages of using OpenSCAD.
The first is related to the parametric capability of programming-based CAD.
P17, for instance, discussed his work in a laboratory making prototypes \textit{`` it was useful to have this programmatic base to create arbitrary variations of similar things''}.
Participants found it helpful to define complex geometries through mathematical definitions instead of storing high volumes of data when having geometric information, as happens in direct modeling.
P2 worked with robotics and talked about his needs \textit{`` I want compact shape descriptions that generate highly complex geometries (\dots) we don't want to store all the geometry with triangles''}.
Programming-based CAD also helps to generalize models better.
For instance, P2 mentioned that resizing a robotic articulated arm involves more than just geometric scaling.
The operation may require adding another articulated section, and describing this behavior in direct manipulation is difficult.
Moreover, programming features such as abstraction allow participants to re-use work.
Further, participants found it convenient that Open-SCAD is open-source, runs on all major operating systems, and is the most popular programming-based application with community support. P12 mentioned \textit{`` It's also the most popular program. If you find a problem, someone else already had it and you can just copy the code or see a different approach''}.

However, some participants (\textit{n = 7}) identified liabilities of using OpenSCAD. 
Despite the support community, they feel that the development of the application is slow. 
P4 mentioned trying to contribute to the project on GitHub, but multiple pull requests have been on hold for a long time without being integrated into the application. 
Moreover, they found that the available features are too basic and the rendering time of complex models inconveniently long. 

\subsubsection{Programmer mindset}

All the participants except P18 (Table \ref{tab:demographics}) had a programming background and found the 3D modeling programming-based paradigm convenient.
P19 expressed \textit{`` I discovered OpenSCAD. I was like, "Oh, hey, this is just models and software" (\dots) It was beneficial to be able to develop parametric models in code, which was part of my skill set.''}.
Programmatic interfaces are a good entry door to 3D printing for this population.

Participants (\textit{n = 13}) also mentioned that OpenSCAD also fits their mathematically oriented mindset.
Interestingly, P12 and P15 expressed that despite having a programmer mindset, one of their problems was their lack of math skills.
P12 said \textit{`` I'd like to say I'm an expert in OpenSCAD (\dots) The one thing that bugs me down is the math. I always get stuck on that''}.

\subsubsection{Experience with direct manipulation programs}

Six participants commented that the direct manipulation paradigm did not work with their mindset.
P11 said \textit{`` I'm not a visual guy, not really an artist. I can imagine what I want to do and write it down without a preview''}.

Interestingly, some participants (\textit{n = 4}) thought that they would inevitably need to learn direct manipulation applications due to the perceived limitations of programming-based ones.
However, some of them have succeeded without learning a new application. P5 commented \textit{`` I've always said to myself, I'll learn AutoCAD when I need it, and so far I haven't needed it.''}

\subsection{Design}\label{sec::theme2_design}

We discussed several aspects of the design process and how it relates to the other stages in the fabrication process.

\begin{table*}[thb]
\Description{The table depicts the structure of subthemes of the theme Design. From the left to the right, the table stores subthemes with their names and the number of participants who commented on them. In the left column are the most specific subthemes. Moving to the right, subthemes are grouped in mode general subthemes until reaching the theme in the most right part of the table. Themes are blue colored with different intensities proportional to the number of participants that commented on the subtheme or theme.}
\caption{Structure of theme \textit{Design}. Color intensity is proportional to the number of interviews coded with codes of the theme and subthemes.}
\label{tab:Theme2}
\centering
\includegraphics[trim={1.8cm 6.4cm 2.cm 1.8cm}, clip,width=.98\textwidth]{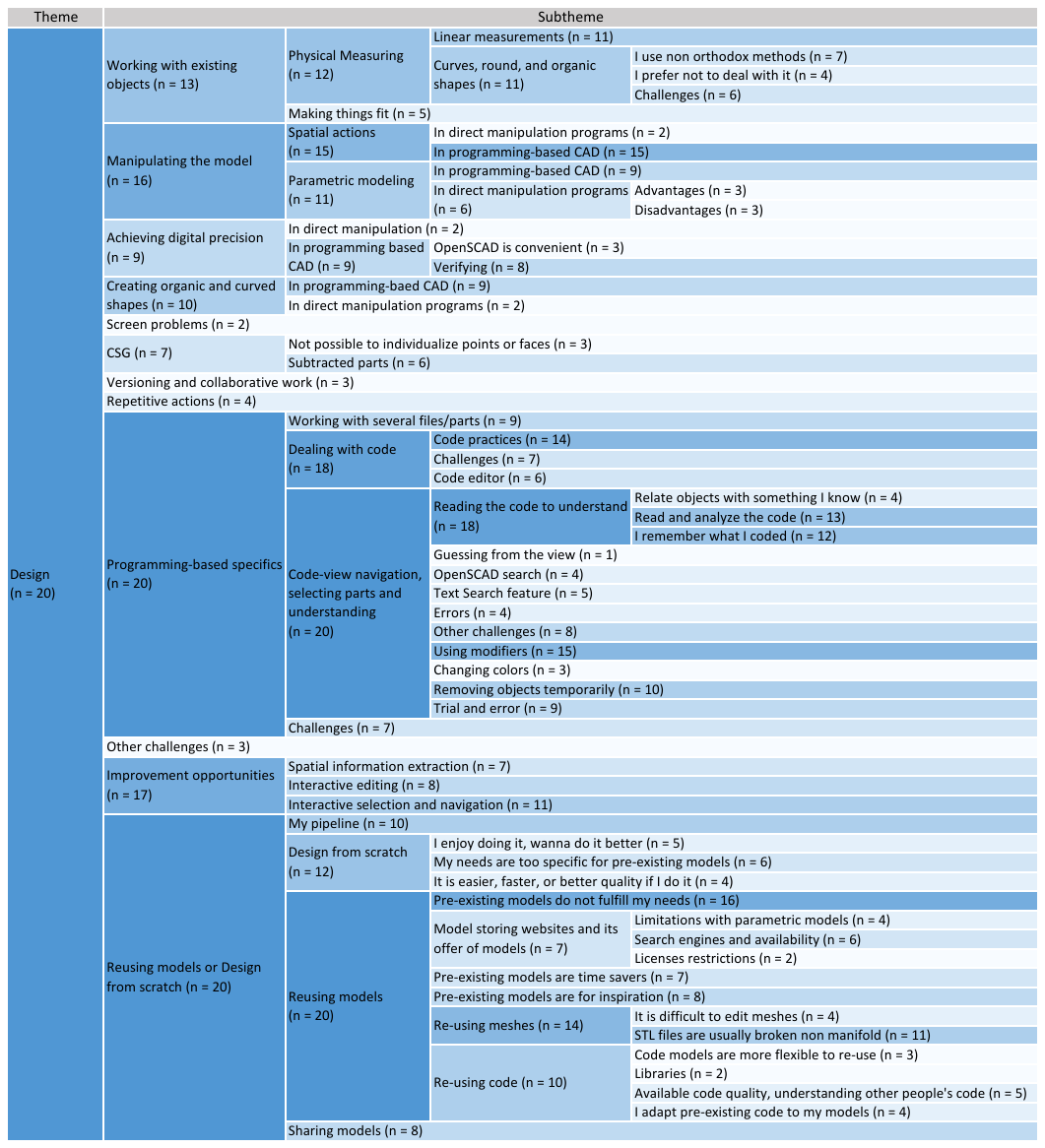}
\vspace{5mm}
\end{table*}

\subsubsection{Working with existing objects}

Often, makers fabricate objects that will interact with other objects, such as in the case of repairing or augmenting an object.
Participants shared their experiences in such cases.

\paragraph{Linear measuring}
The preferred measurement tool for linear measurements is the digital caliper.
However, two participants stated that the task of measuring increases uncertainty and leads to more iterations.
P8 mentioned \textit{`` If I was better at taking measurements, I would go through fewer iterations, and my first print would probably be closer to what I want''}. Moreover, calipers have size limitations, according to two participants. P14 commented \textit{`` calipers don't go big enough to measure a lot of this stuff''}.

\paragraph{Measuring organic shapes and curves.}

In addition, measuring organic or curved shapes is complex (\textit{n = 11}).
For instance, P15 commented on their work on repairing parts: \textit{`` it's rectangular and then there's a curve. I had to print it like 15 times to get that right.''}.
To deal with it, seven participants reported using creative solutions.
For example, P12, P14, and P20 have used cameras or scanners to get the outline of a shape and use it later in the design by transforming it into an SVG or by measuring the outline and approximating the curves.
P14 commented \textit{`` I photocopied the object, I put it down on a photocopier, so I could get a picture of the rim of the profile, from which I could make measurements of it. Then I had to run an optimization program in a spreadsheet to figure out how everything fits together, how all the curves match, and what angles join each other\dots''}. P14 also reported using photogrammetry with no good results.\textit{`` I tried photogrammetry to make a 3D model of this; it just doesn't work if you have shiny or transparent surfaces. I got a nice model but also sort of a cloud of nondescript points because of all the reflections on the surface''}.
Other participants have tried to measure some points to interpolate by guessing in a trial-and-error strategy. P12 mentioned using a contour gauge to approximate curves and
P14 said that at some point, he would hold the physical object in front of the screen to see if the object to print would match.

Some participants (\textit{n = 4}) prefer to avoid organic shapes and curves. P20 commented that \textit{`` I try to avoid them (curves and organic shapes) in my designs. If I'm just talking about rounding corners, I pick up a set of radius measurement tools so I can get the correct size of rounded edges. Beyond that, it's trial and error.''}

\paragraph{Using digital replicas.}
It is easy to miss context elements when fabricating objects interacting with other things (\textit{n = 5}).
Makers cannot imagine every aspect of the physical objects in the 3D view to see if the design will satisfy their needs.
P9 explained \textit{`` this piece is a cover for an emergency button that you need to screw in manually. I did not think about it in my first iteration, so I could not access the screw hole and had to repeat it, adding a small hole''}. Some participants create a digital representation of the physical object to have a reference to work with, making them more confident about the design decisions.
For instance, P11 used an STL model in a project for his phone: \textit{`` I've used a model of my phone for that. I just use an STL of the phone and design around it''}.

\subsubsection{Spatial transformations}\label{them:cat:spatTrans}

The keystone modeling action is manipulating objects' position, orientation, and size. Participants discussed the difficulties they usually face when performing this task in OpenSCAD and compared it with direct manipulation applications.

Some participants (\textit{n = 2}) acknowledge how easy this task is in direct manipulation applications.
P19 commented: \textit{`` With Fusion360 or TinkerCAD (it is handy) to make a cylinder of the right dimension in the right place, basically drag and drop and get it where it belongs''}.
On the contrary, the same task in programming-based CAD is reported to be very difficult (\textit{n = 15}).
Trying to place an element in the right place can be challenging, as commented by P7 \textit{`` Most times, my difficulty is not drawing but where it should be drawing. Like I aim to draw a sphere here and OpenSCAD put it over there, and I am like "why?"}.

It seems to be challenging to relate the transformation parameters of a code statement with the spatial coordinates in the view without visual cues.
For instance, if a translation is applied with 10 units in the second parameter on a sphere, it is not easy to predict the direction the sphere will move in the view.
The camera view can be in a position and orientation that may make it hard to understand where the axes are located and there is not enough visual help for the user.
OpenSCAD view has a widget representing the canonical \textit{x}, \textit{y}, and \textit{z} axes directions.
However, users often need to locate coordinate systems different from canonical ones.
The position and orientation of objects are typically the result of multiple nested spatial transformations.
Each transformation has a scope where the relative coordinates system does not match with the canonical one, so the widgets are useless.
P15 said \textit{`` if you are creating some volume (\dots) it is difficult to predict, with all the operation, where they land and what precise coordinates would be''}.

Participants also found dealing with translations and rotations of the same object challenging.
These spatial transformations are not commutative.
In other words, applying a \texttt{translate} after a \texttt{rotate} would not give the same result if the commands are executed in the opposite order.
P19 commented \textit{`` you need to think about how you want to translate and rotate it before you can even get it to where you want. (Otherwise) You might find that you get your translation operations out of order, and all of a sudden, you're in the wrong place''}. To deal with this, participants use a trial-and-error technique.
P6 mentioned \textit{`` If you would ask me right now, to rotate this in a certain direction, I would not, without testing, be able to tell you what combination of the three parameters I need to get it in the direction I want''}.
P9 commented on having a specific order for transformations: \textit{`` I rotate first and translate later. It is easier because when I translate before rotation, sometimes the rotation center is not the same as the object's center.''}
The participants proposed another strategy implementing position and orientation checkpoints.
They correctly generated the transformation required for placing objects where they belong.
The elements were then designed in the canonical coordinate system, and after completion, the participants applied the previously calculated transformation.
P17 commented \textit{`` I remember doing like a checkpoint where I start. I make sure that everything starts from this origin point. Then, when you add in multiple modules, make sure that they're centered around so you don't have to keep track of all the different systems.''}.
Similarly, P16 commented that he created modules solely to place elements in a location and orientation of interest.
Some participants mentioned avoiding having several layers of transformations because it becomes unmanageable.

In addition, it was deemed challenging to calculate the appropriate parameter values for the spatial transformations.
As objects' position and orientation are built upon multiple transformations and involve the sizes of other objects, the coordinates system changes constantly, and participants must mathematically derive parameters' values of spatial transformations.
Depending on the previous spatial transformations, the relative coordinate system varies.
P14 commented \textit{`` \dots I have to painfully mathematically calculate where that plane is in space, its slope, and where its normal vector is, and then get the surface on there\dots''}.

\subsubsection{Parametric modeling}

The creation and use of parametric models were reported as valuable for the participants (\textit{n = 11}).
For instance, P14 stated that direct modeling is not enough for serious developments \textit{`` TinkerCAD is easy and fun, but not useful for parametric modeling or anything serious.''}.
Further, they value the possibility of revisiting their steps.
P7 said \textit{`` because everything is written down \dots if I pick up something one year later, I'll remember exactly what I did.''}.
Moreover, they made clear differences between creating a parametric model in direct manipulation software and OpenSCAD.

Six participants shared their thoughts about creating parametric models through constraints in direct manipulation applications such as FreeCAD and Fusion360.
The ability to select objects of interest directly from the view is perceived as practical, as mentioned by P14~\textit{`` you can grab a vertex and snap it to some other location and not have to worry about numbers and measurements so much, but you can make things fit just by moving things around and snapping things to grid points or to other vertices, I find that's very nice.''}.
However, the constraints management in such programs was perceived as difficult.
Participants found tracking constraints challenging because they are represented as a list in a panel different from the view. In consequence, it is easy to get into a model with several constraints that are hard to edit or that end up being overconstrained. P9 commented \textit{`` I try to avoid using constraints normally because when I've tried, I've ended up stuck, I have too many constraints and keep receiving the "Overconstrained" error message, and it is not easy to fix it for me. I do not know how to know which constraint is the problem''}.

Some participants (\textit{n = 9}) commented on the way of creating parametric models by defining object properties through parameters and variables in OpenSCAD.
All agreed on the importance of generalizing model behaviors through the use of variables and avoiding the definition of object properties with hard-coded numbers.
P6 said \textit{`` \dots everything is based on making the outer frame as a combination of parameters, so the parameters can change and then the model still works''}.
However, participants mentioned that defining everything in terms of variables can be exhausting, so sometimes, they use variables only when they anticipate that a certain property will need to change.
Moreover, working out the mathematical expressions for creating parametric models is perceived as very challenging.
P6 expressed \textit{`` I sometimes have a harder time doing the math (to define object's properties), using all the combinations of variables where they should be''}.

\subsubsection{Achieving digital precision}

One primary need of some participants (\textit{n = 9}) when printing is to achieve precision.
They perceive that OpenSCAD allows them to achieve accuracy in easier ways by explicitly expressing the sizes and dimensions of every placed part.
P1 commented \textit{`` from Blender I could not make things precise (\dots) I use CAD programs for electronic devices, from there I get measurements. In OpenSCAD, it is very easy to be precise with this. I only need to make a box of this size or that size \dots''}. P12 acknowledges that some direct manipulation programs allow one to express precise measures but it is not as clear and flexible.

Paradoxically, while users feel a sense of precision by being able to describe position and sizes explicitly, eight participants complained about the lack of means to check dimensions in the OpenSCAD view. P19 mentioned \textit{`` I had a really tough time taking measurements and showing measurements visually as part of the model.''}
As mentioned, spatial transformation involves several nested operations and variables, so verification is important.
P14 said \textit{`` I wrote all these formulas, but did the resulting piece have the correct size and location?''}.
To deal with this, participants use \texttt{echo} operations to print the expression results in a console (P4 \textit{`` I could put some echos to verify those formulas, but It would be much better measure on the screen.''}) or visually inspect parts on the view by emulating a ruler (P11 \textit{`` \dots I put a cube of the right size next to it and just visually inspect if they are the same height''})

\subsubsection{Designing curves and organic shapes}

Nine participants said that, in general, they feel that OpenSCAD is not a friendly application for creating nonstructured curves and organic shapes that are difficult to define using mathematical expressions.
For instance, creating smooth corners was reported as \textit{difficult} and \textit{painfull}.
Participants commented: P4 \textit{`` (in OpenSCAD) making rounded edges is a pain''}; P13 \textit{`` It is just easier for the printer to 3D print something that's rounded (\dots) But designing it, that is really super hard to do in OpenSCAD''}. Even using prebuilt OpenSCAD functions for this purpose is hard, such as \texttt{Minkowski} function.
P16 and P1 mentioned \textit{`` (the most difficult in OpenSCAD is) figuring out how to do rounded edges and fillets and chamfers without using Minkowski because it is too time-consuming to render''} and \textit{`` \dots making smooth corners is possible, I use the Minkowski tool but the dimensions changes, it is inconvenient.''}.
However, three participants said that when the shape or curve can be defined mathematically, OpenSCAD is convenient for designing it.
P10 discussed a project where he defined a Bézier curve that changed parametrically.
Although the mathematical expression was hard to create, it would work better to have a parametric design because the curve can be expressed in code.

On the other hand, this task seems significantly easier in B-rep direct manipulation programs.
P6 mentioned \textit{`` in Fusion360 if I want to have a squared box and I want to have a nice rounded corner, I just can click on both faces and I have a nice rounded edge''}.
This difference in difficulty between programming-based and direct manipulation applications in this seemingly simple task creates frustration for the participants.
P4 expressed \textit{`` What makes it hard is that you can not point to a specific edge or corner and make it round. I don't like that''}

\subsubsection{Dealing with CSG}

Some participants (\textit{n = 7}) mentioned limitations inherent in the CSG representation.
First, four participants said it is hard to verify the result when using the \texttt{difference} and \texttt{intersect} operations that remove volume.
The removed parts are not visible, and verifying the correctness of the operations seems problematic.
P4 commented \textit{`` When you do a subtraction, it is hard to figure out if you are doing it right, if the subtracted piece is in the right spot, and if it has the right shape''}.
One mechanism to deal with this difficulty is using OpenSCAD \emph{modifiers} (see section \ref{sec::os}). However, modifiers require a trial-and-error approach, which is still time-consuming.
P16 expressed \textit{`` (I use modifiers) with the invisible things, the things I'm subtracting. I'd said it's an easy way to figure it out, but it always takes some time''}.

Second, three participants expressed frustration that they were unable to individualize points or faces from the volumes as it is possible with B-rep.
For instance, P4 mentioned it was hard not to be able to point a corner and round it from the view. A similar problem occurs when, for instance, the model requires two cubes to touch in one face.
By placing one box at the end of the other, there is no guarantee that the objects are overlapping.
CSG does not represent faces or vertices information but abstract definitions.
Hence, users can only define two boxes with coincident faces by correctly defining their positions and sizes.
Thus, it seems to be common practice to add small offsets to ensure overlapping as mentioned P13 \textit{`` You can't have coincident faces; they have to go past each other. Every time you put one thing on top of another, you have to add those overlaps''}.

\subsubsection{Versioning and collaborative work}

Few participants (\textit{n = 3}) highly valued the flexibility of code to use repositories such as GitHub for versioning and collaborative work.
Although it is possible with direct manipulation programs such as FreeCAD, it is not that convenient, as expressed by P3 \textit{`` I especially like how well OpenSCAD checks in the GitHub repository \dots this is a very collaborative project, and you can check in Fusion360 in GitHub but it just does not work. The repository becomes huge very fast. But OpenSCAD is text-based, making it very well suited to the collaborative environments we are already using for the source code.''}

\subsubsection{Specifics of programming-based CAD}

We discussed with the participants the advantages and limitations of programming-based CAD applications in general and specifically in OpenSCAD.

\paragraph{Working with several files/parts}

It is normal to break down the model into parts when having complex models.
In addition to facilitating the design process, it is an alternative when the part has to be printed in several parts.
This scenario presented some difficulties for the participants (\textit{n = 9}).
OpenSCAD does not have any option to define parts in the design, so the participants can export parts individually.
One solution mentioned by the participants was to create the design and later apply a \texttt{difference} and \texttt{intersection} with a cube.
First, the participant would apply the \texttt{difference} and remove half of the model to export it to an STL file for printing.
Later, the participant would use the same cube and apply an \texttt{intersection}, removing the second half to export and print.
Once both parts are printed, they will be assembled.
The result would not make parts easy to connect.
Therefore, other participants used some libraries to split the model into parts to better assemble them.

In other cases, applying \texttt{difference} and \texttt{intersection} is not enough when the model is complex.
Thus, participants create a file for each part.
This allows them to isolate each part and focus their efforts without being distracted by the other parts.
Nevertheless, the process of validating all the parts together is very difficult.
For instance, P5 shared with us a model of a GoPro camera gimbal with many parts that form an articulated arm.
He imported all the parts into the same project when he wanted to verify how it would look together.
However, when modeling individually, all pieces are placed in the center.
Thus, when importing all parts together, it is not easy to place them in the correct position and orientation only to verify the result.

\paragraph{Dealing with code}

Eighteen participants discussed issues and experiences managing code in OpenSCAD.
In general, participants try to keep good coding practices to make their models easy to understand.
They mentioned the importance of commenting on code, avoiding very long modules, organizing the code, and using expressive and telling names for variables and modules.
However, they acknowledge that using expressive names for all variables is impossible.
Moreover, using the expressiveness of the language is not always useful.
For instance, P15 explained his model, which was extensively documented in his mother tongue, Slovakian.

Some participants (\textit{n = 7}) discussed the challenges they face when dealing with code.
According to them, problems include difficulties in easily finding parts in the code, keeping track of variables on complex models, and refactoring code.
Moreover, we observed another challenge in the hands-on exercise.
When participants were looking for a code statement of a particular part, they would analyze and perform tests on the code that was not being evaluated.
For example, elements created inside a conditional structure that was not evaluated.
OpenSCAD does not warn users about this situation, and they realize this after spending some time analyzing the code.

Some participants (\textit{n = 6}) stressed that OpenSCAD code editor is basic and lacks more advanced code editor features, as commented by P6 \textit{`` OpenSCAD really lacks richness in helpers how to write code, there is no autocompletion and that kind of stuff''}.

\paragraph{Code-view navigation}

In programming-based CAD, the description of the model (\textit{i.e.} code editor) is separated from the model visualization (\textit{i.e.} viewer).
Thus, users must constantly switch between the code where they edit the model and the view where they validate the result of the modifications.
As a result, navigating the model and making edits can be difficult \cite{yeh_craftml_2018}.
Some participants opted to use Visual Studio Code (VS Code) \cite{microsoft_visual_2023} instead of using OpenSCAD code editor.
They stated that VS Code allows for the installation of OpenSCAD plugins that streamline the coding process.
The participants used VS Code to modify their code files and accessed a separate window within OpenSCAD to review the output.
We observed the behavior of the participants in the hands-on exercise and discussed with them the challenges related to these tasks.

We identified a three-step search pattern when looking for code statements that create specific parts in the view.
First, they would try to identify the block of code where the target part could be defined.
Then, they would study the code to confirm the selected code statement logically.
Finally, they would seek a visual confirmation in the view.

Participants had five strategies for trying to locate the code statement based on the view: rely on their memory, link the part to a variable and search the variable, guess how the part should be created and look for the pattern, follow the comments, and using OpenSCAD search feature (see section \ref{sec::os}).
As participants worked with their own models, some participants (\textit{n = 12}) tried to remember how the model was built and relate it to their normal way of structuring the code.
For instance, when P7 tried to find the code statement of a part, they said \textit{`` I know how the hex array (main frame of the model) is organized in the first play, so I would go here (scrolls the code until finding the hex\_array module), ok here it is''}. The second strategy was to link the target part with the variables and use a text search feature, as P17 commented \textit{`` I would assume that it's related to this variable''} before searching for the occurrences of the variable.
This strategy did not work well when the model repeatedly used the variable the participant picked to locate the target.
The large number of occurrences made it difficult to study all of them to decide which code statement was correct.
Moreover, OpenSCAD code editor features are basic and do not provide visual cues to help developers understand the code (\textit{e.g.} highlight calls in the scroll bar of a selected variable or jump to the definition of a selected module by clicking on the module call).
Participants who use VS Code could easily follow the places in the code where a variable was used. In some occasions, participants using VS Code made mistakes by editing the code of a file different from the file OpenSCAD was rendering.
They edited the code and did not see any change in the view until they realized the problem after some time.
The third strategy was to try to think about how the selected part should have been created in the code and look for that pattern in the text editor.
For example, the target element for P6 and P2 was a hole, so they started looking for a \texttt{difference} code statement.
In the fourth strategy, when the model was well documented, participants used comments to understand the structure of the code and find the correct statement.
Finally, participants could locate the code statement with the OpenSCAD search feature.
Only three participants knew about these features and mentioned not using them normally.
When seeking the code statement, participants always read the code to understand it and confirm that it was the target statement.

After the participants thought they had located the target code statement, they normally sought visual confirmation.
The strategies used to confirm were removing the code statements and verifying missing objects, changing the parameter values of the code statement and verifying changes in the object's properties, using a \texttt{color} operations to highlight the object, and using OpenSCAD modifiers as a debugging task.
We could identify some challenges when performing these strategies.
Removing code statements can break the syntax of the code.
For example, when a \texttt{translate} statement is written without opening and closing brackets, the transformation is applied only to the next code statement.
The system will report an error if the code statement is temporarily removed and the next statement is not an object (\textit{e.g.} variable definition).
Also, using \texttt{color} operation does not work if the object is already inside a \texttt{color} scope. The system will override the statements, prioritizing the statement placed higher in the tree of operations.
Participants frequently use modifiers for visual inspection but they often forget the correct syntax and characters to use.
Moreover, modifiers do not work when they are used on 2D elements or code statements that are not being used (\textit{e.g.} not evaluated conditional).
In any case, all participants used a trial-and-error technique and required reading and understanding of the code.

Some participants (\textit{n = 8}) mentioned some challenges they identified after the hands-on exercise.
They acknowledge that reading and understanding the code is difficult.
It becomes more challenging in complex models with long scripts, highly decoupled with several module calls, and with several parameters.
They also mentioned that finding code based on the view is a repetitive, hard, and mentally demanding task.

\paragraph{Other programming-based CAD challenges}

Eight participants mentioned some difficulties they identified when designing with OpenSCAD.
We observed that participants are often surprised by the changes performed on the models because it is impossible to anticipate the edit's result confidently.
Programming-based applications do not provide a transition between two states of the code.
Consequently, understanding the impact of the changes made to the code is not always easy or obvious, as commented by P2 \textit{`` One of the difficulties is that you don't have immediate feedback when you change something as a result of the screen; there are always delays.''}.
Other challenges were related to difficulties in managing text elements in the view and confusion with 
OpenSCAD units.

\subsubsection{How to improve OpenSCAD}

Participants (\textit{n = 14}) shared ideas about features that they considered would help their modeling experience.
One key idea was the capability to easily identify parts of the code based on the view and vice versa.
Participants thought that there could be features to help with this task.
In particular, they would like to have a visual cue that helps them locate the code based on interactions with the view.
P13 commented \textit{`` If I could point at something in OpenSCAD and tell me where this is (in the code editor), that would be a huge help. Especially if you have complex geometry, it's just super hard to figure out just where you are.''}.

Participants also commented on how to facilitate the task of applying spatial transformations.
First, they mentioned the need to be able to measure distances in the view.
They also commented that it would be very convenient to be able to extract spatial coordinates of objects directly from the view.
P14 commented that  \textit{`` There are cases where I want to know all those coordinates, but OpenSCAD doesn't give that to you. I would like it to tell me what the bounding boxes of my model are. It doesn't have to show me on the screen, but at least when I render it in the output area, it would tell me the bounding box and the center of mass''}.
Moreover, seven participants indicated that they would like the system to assist them in retrieving the spatial coordinates in terms of the existing variables for parametric models by interacting with the view.
P3 said  \textit{`` (I would like) if I could click on this point (in the view) and OpenSCAD would automatically create an expression using the variables to describe that point (\dots)not hardcoded numbers because it is not useful anymore that way''}.
Further, in addition to working out the expressions, they would like the system to allow the creation of constraints directly from the view.
P5 commented  \textit{`` So being able to simply point at an object to say, I want to attach this face of this item to that point there and for it to be able to calculate that distance would be such a timesaver''}.

Finally, participants mentioned the need to facilitate some recurrent tasks for 3D printing.
Specifically, they mentioned that they would like to have more elaborate libraries to perform usual actions, such as creating chamfers.

\subsubsection{Designing from scratch or re-using models}

We discussed re-using models from websites such as Thingiverse, Printables, or Github with the participants.
Half of the participants (\textit{n = 10}) expressed that it is normal to check pre-existing models from websites before starting to design one from scratch.
However, the reasons for that differ according to several aspects.
For instance, P8 said that he usually starts by searching for models in Google when it is something complicated.
This is probably because P8 has a short experience with OpenSCAD and 3D printing.
Nine participants mentioned that if they needed to print something very popular, they would definitely go to check other people's solutions first.
P20 said  \textit{`` if it's something that's really popular that I absolutely know that there would be models for (\dots) I'll just go get one rather than design yet another one''}.

Nevertheless, opinions on preferences between re-using models (and what using them for) or designing from scratch are divided.

\paragraph{Designing from scratch}

Six participants prefer to design from scratch because they have specific needs and look for very customized models. 
P4 commented  \textit{`` I can design to fit my own setup \dots for instance, my screwdriver holder, all the holes are a different size to fit my particular set of screwdrivers so that they fit snug\dots''}.
Other participants (\textit{n = 5}) expressed their satisfaction in challenging themselves to build models of good quality on their own, as mentioned by P11  \textit{`` I just like the process of designing and printing things and be ready to say, I've done this by my own.''}. Furthermore, four participants said that starting from scratch is easier, faster, and has better quality results.

\paragraph{Re-using pre-existing models}

We discussed with participants their thoughts about using pre-existing models.
We asked them if they use them, their motivations, limitations, and the model format they prefer to look for.
Most of the participants (\textit{n = 16}) reported that pre-existing models do not meet their needs.
Even when several models exist, they usually do not exactly fit participants' needs.
And even if the model is parametric, available parameters rarely cover the modification participants want.
Moreover, some models are too complex, with several parameters adding significant complexity to the printing process, so they are discouraged from using them.In

 addition, participants perceive several difficulties with the available model-storing websites.
To start, six participants commented that searching on these websites is challenging.
The naming system is deficient, so it is difficult to find useful models.
Moreover, dealing with licenses can be problematic.
Some participants use 3D printing for commercial applications, so using public models is not ideal for them.
Participants also commented on the \textit{Customizer} application from Thingiverse.
Although they find it useful, they also mentioned that it is very limited because authors can only upload models with one file, without the possibility of making references to other files or libraries.

However, the participants also acknowledge the potential of such websites.
Six participants mentioned that pre-existing models can be time savers.
They said that for printing popular objects, and for people with little experience in design, it is a good alternative.
Some participants (\textit{n = 8}) used pre-existing models as inspiration.
These websites present alternatives for participant's projects in progress where they collect ideas.
Some of them mentioned being surprised to see things they never thought possible to do with 3D printing.

Although participants did not use pre-existing models to print them directly, some have incorporated or edited pre-existing designs for their projects.
Fourteen participants commented on difficulties in editing STL files in direct manipulation and programming-based applications.
All of them agreed that finding manifold and printable geometries is rare.
In most cases, they needed to fix broken geometries before being able to use them, which was reported to be \textit{`` difficult''} and \textit{``painful''}.
Participants preferred to have a pre-existing model \emph{with code} instead.
The offer of models with source code is more limited and quality varies significantly.
Often, available models are not coded following good practices of programming, such as adding documentation, so reading and understanding the code is very hard.
Some participants mentioned the importance of good-quality code if it is meant to be shared as mentioned P5 \textit{`` One thing is designing for yourself, and other designing for sharing''}
This is not always the case and depends on the author.\enlargethispage{12pt}

\paragraph{Sharing models}
Some participants (\textit{n = 8}) had profiles on websites such as Thingiverse and Printables, where they shared their designs.
They manifested that they enjoyed sharing their models, contributing, and seeing other people using and commenting on their models. Three participants said that although they liked sharing, they did not want to spend more time adjusting their models. Unfortunately, the models need to be adjusted to share parametric models in applications like Thingiverse Customizer. For instance, Customizer only accepts one-file models. P5 commented that he would like to share more of his models, but he often re-uses his own libraries that cannot be uploaded to the website.
P6, on the other hand, said that he would like to customize the parameters to provide a better experience using sliders instead of input boxes, but would not spend time learning how to do it.
Also, some participants mentioned that websites are not controlled, and they had seen users taking models from authors and selling them on other websites.

\begin{table*}[htp]
\Description{The table depicts the structure of subthemes of the theme Printing. From the left to the right, the table stores subthemes with their names and the number of participants who commented on them. In the most left, the most specific subthemes. Moving to the right, subthemes are grouped in mode general subthemes until reaching the theme in the most right part of the table. Themes are purple colored with different intensities proportional to the number of participants that commented on the subtheme or theme.}
\caption{Structure of theme \textit{Printing}. Color intensity is proportional to the number of interviews coded with codes of the theme and subthemes.}
\label{tab:Theme3}
\centering
\includegraphics[trim={1.8cm 18.5cm 2.8cm 1.8cm}, clip, width=.98\textwidth]{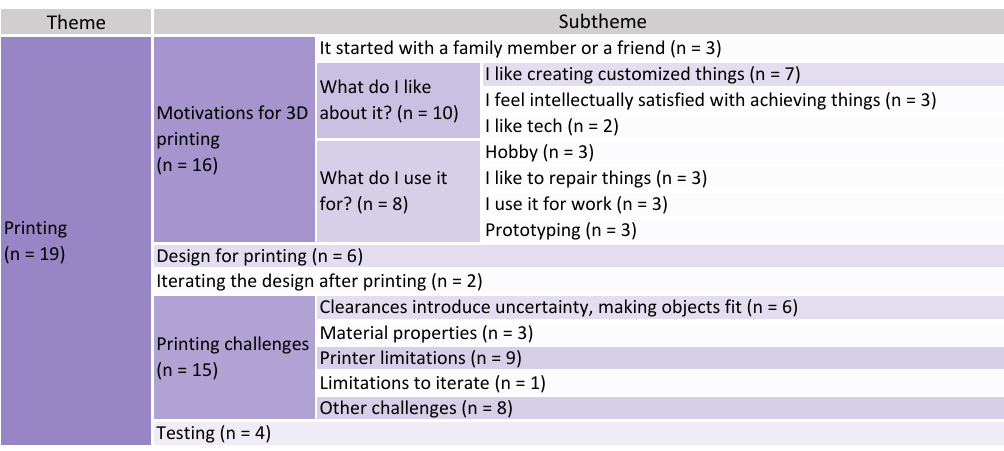}
\end{table*}

\subsection{Fabrication}\label{sec::theme3_fabrication}
Some of the questions in the interview were intended to identify challenges in the fabrication process.
We discuss our findings related to motivations and challenges.

\subsubsection{Motivations for 3D printing}

Participants use 3D printing for prototyping, for work, for repairing other objects, and as a hobby.
They discussed what they liked about 3D printing.
Five participants said that they were interested in technology and applications so it was almost natural for them to try 3D printing.
Other participants mentioned that they have an \textit{`` intellectual satisfaction''} when they fabricate complex customizable things successfully.
In addition, seven participants mentioned enjoying the fabrication process and especially having the physical result.

\subsubsection{Design for printing}

Three participants mentioned that designing and designing for printing are two different ideas.
As explained by P11 \textit{`` Surely you can design everything in CAD, but having prints with many overhangs or supports is impossible. You need to design your models with printing in mind''}.They also commented that it is easy to find very complex models that, in practice, are not printable.
When rendering a model in OpenSCAD, users can use a rapid preview or a more realistic rendering option.
P3 commented that these differences add uncertainty to the design process because the way both rendering options process the information can lead the designer to errors.

\subsubsection{Printing challenges}

Participants (\textit{n = 15}) discussed different problems encountered with 3D printing.
Very often, they expressed that tolerances and clearances are factors that introduce high uncertainty on the success of the printed object.
3D printing does not precisely reproduce the dimensions and sizes in the digital design.
P10 commented \textit{`` I usually have to print multiple times to get the clearances correct, especially if there are moving parts. It usually takes several times to get the tolerances right''}.

Another challenge is the difficulty of taking into account the material property in the design process.
Different materials have different behaviors that are difficult to include in the design because OpenSCAD does not support this information.
For instance, how to design taking into account weak points in the design.
P14 commented that\textit{`` most difficult in designing for 3D printing is making sure you're accounting for the anisotropic strength properties of the material \dots anything you print is weaker along layer lines, for instance''}.

Nine participants considered that knowing the printer they use is a key factor in minimizing uncertainty.
They mentioned that hardware in 3D printing is sensitive to failure and requires good skills to set up for success, as commented by P19 \textit{`` if it's your own printer, you have a better sense of what to expect and you have agency over the state of your machine''}.

Other limitations were discussed.
When the model is large, there is less flexibility to iterate due to the amount of material required.
Moreover, participants mentioned that sometimes, putting models in a correct orientation for better printing is not trivial.
Furthermore, some participants (\textit{n = 4}) mentioned that printing with supports can be tedious, so they prefer to avoid it when possible.

\subsubsection{Testing}

Testing was important to avoid false printings and waste of material.
Participants (\textit{n = 4}) acknowledge that their only validation strategy was test prints.
P13 and P16 said they would print layers to verify clearances before printing the entire piece.
Although test prints are time-consuming, they were unable to find another testing mechanism to anticipate what the result of printing would look like.

\section{Discussion}

We are interested in understanding the motivations of programming-based CAD users. Moreover, we aim to study their challenges and limitations to discuss opportunities for HCI research.

\subsection{Programming-based CAD users in 3D printing}
Findings in programming-based CAD user preferences (section \ref{sec::theme1_profile}) allowed us to identify two types of motivations when users 3D print.
The utilitarian and the enthusiastic. Users with utilitarian motivation use 3D printing to solve problems pragmatically. They are more flexible in design decisions when there are limitations or a lack of motivation.
For instance, participants expressed not liking re-using pre-existing models, but due to the time limitations, for example, P19 found pre-existing models a very convenient solution.
He expressed \textit{`` As a father with a full-time job, it's difficult to sit down and develop a model (\dots) So oftentimes, I'll look for existing models, and if they work, I go with them because it's often easier''}.
Similar situations occurred when printing common objects that participants felt they could find on the Internet.

In contrast, users with enthusiastic motivation invest time and energy in achieving neat and highly customized objects.
Participants used very creative methods and could iterate several times to achieve a satisfactory result.
P4 shared how a small model for a screwdriver hole changed over 5 iterations to achieve a satisfactory result.
P14 had experimented with different creative solutions to capture the outline of curved objects he wanted to repair.
When having an enthusiastic motivation, users are more likely to design from scratch rather than use pre-existing STL models with questionable quality and printability, or coded models that would require time and effort to understand.
Moreover, some participants were moved by the feeling of pride of \textit{`` intellectual satisfaction''} when achieving a result. As commented by P16 \textit{`` I like OpenSCAD because it is very functional, not procedural. That's an intellectual like \dots functional languages (like OpenSCAD) are intellectually satisfying for me.''}

Furthermore, we identified that the fabrication process involved stages that users can or cannot enjoy.
P7 mentioned that \textit{`` (3D printing) It's actually three hobbies in one, which is why a lot of people find it very overwhelming (\dots) There's the modeling, setting and improving the 3D printer, and the actual making of the 3D prints''}.
Indeed, we could observe different enjoyment from the participants at the different stages of the process.
For instance, P14 described a project that took him weeks to create a highly customizable model for pliers.
On the other hand, P16 explained a workflow he developed to create objects with multiple colors on a single-color printer by playing with the printer settings and the design.
Moreover, P8 said that, although he enjoyed the design process, what he liked was the result.
Consequently, he rushed other parts despite the errors he could make, such as taking measurements with care.
\textit{`` I think like to go to the good part, I want to get to the fun, having the thing in my hand.''}
Previous studies \cite{hudson_understanding_2016} reported that novices were easily frustrated designing objects for 3D printing, probably because their motivations were to get the objects, not to design them.\enlargethispage{12pt}

\subsection{Programming-based CAD design challenges}
The interviews reveal challenges that programming-based CAD users face (section \ref{sec::theme2_design}), representing a research opportunity for the HCI community.

Programming-based CAD inputs are coded instructions that the machine uses to create a visual representation where the user can verify the result and perform edits in the code if required.
Therefore, the user verifies in one space and edits in another in an intensive and iterative exercise.
Consequently, users are forced to understand the connections between the code and the rendered model, but current tools barely try to help users in this task.

\paragraph{Navigability}
Our hands-on exercise revealed that users use several strategies to identify code statements responsible for creating specific parts based on visual inspection with considerable difficulty.
Most of the participants went through the code to understand it, and in every case, they needed to make edits to seek visual confirmation.
Programming-based applications could facilitate this task by using the implicit connection between the code and the view as some direct manipulation applications do, connecting the view and the history tree.
Applications such as FreeCAD \cite{the_freecad_team_freecad_2022} and Fusion360 \cite{autodesk_inc_fusion_2023} allow users to click on specific parts on the view while the corresponding element in the history tree is highlighted.
OpenSCAD provides a backward search (section \ref{sec::os}) to place a cursor on a code statement based on a selected part in the view.
However, it does not provide visual cues to confirm that the selected part is aimed, and users need to modify the code to seek visual confirmation.
IceSL \cite{lefebvre_icesl_2022} provides better visual cues to highlight code statements responsible for creating parts in the view but does not discriminate between statements involved, so it is impossible to navigate the CSG structure.
Gonzalez \textit{et al.} \cite{gonzalez_introducing_2023} propose a navigation system that helps isolate specific parts with the corresponding code statements with visual cues in OpenSCAD.
Such systems facilitate the understanding of the structure of the code and how it is related to the view without the need to study or modify the code.
However, this interaction technique has limitations related to ambiguity when selecting parts that are visually the same but different in code.
For instance, a cube is modified by a \texttt{translate} or \texttt{scale} statement.
The visual response is the same when selecting any of these two statements, although each performs different tasks.
A navigation system could differentiate these cases, adding visual cues to understand the nature of the statement.
In the previous example, the preview could show an animation of the cube being created and moved or scaled, with the transformation statement indicating the dynamic effect of the code.

\paragraph{Understanding spatial coordinate systems in the code}
Understanding 3D spaces on a 2D screen is a difficult task reported in direct manipulation \cite{hudson_understanding_2016}.
Our interview revealed that this problem can be more difficult in programming-based CAD where the editing space is disconnected from the view, and there is no visual help to relate them.
In other words, users have to mentally imagine the behavior in the view that a spatial transformation will have when stated in code.
Moreover, CSG structures create nested scopes where the coordinate system is relative to the aggregated effect of spatial transformations performed previously.
Participants mentioned the need for trial and error strategies to apply a spatial transformation successfully.
The task becomes more complex when using rotations, which are generally more difficult to understand.
Some participants tried to avoid nested transformations due to difficulty understanding them.
To alleviate these challenges, programming-based CAD applications could incorporate features to enhance users' comprehension of spatial properties during coding.
For instance, some direct manipulation applications integrate \textit{"manipulators" }\cite{jankowski_advances_2015}, such as row-shaped widgets positioned at the center of objects, indicating the relative 'x', 'y', and 'z' axes, as seen in Unity \cite{technologies_unity_2023}.
In programing-based CAD, Gonzalez \textit{et al.} \cite{gonzalez_introducing_2023} system allows the user to select a specific part and places these manipulators in the center of the relative coordinate system of a selected object to visualize it.
This helps to understand the relative system coordinate in the view, but it does not connect it to the meaning in the code.
The correspondence between both spaces can be more explicit through visual cues.
For instance, axis widgets have different colors to distinguish the translation axis, which the code editor could use in the space of the transformation parameters to make this correspondence explicit.
Moreover, by clicking on the widgets, the application could open a text dialog to edit the corresponding parameter directly in the view, as is possible in some direct manipulation applications \cite{autodesk_inc_fusion_2023}.

\paragraph{Defining geometric properties based on pre-existing information}
In CAD design, the geometric properties of an object are closely related to those of other objects.
For example, users may want to place a cube on top of another.
Direct manipulation applications facilitate user interaction through visual aids such as rulers and volume highlighting during overlap, the use of snap effects that guide positioning during drag-and-drop actions, or the use of constraints \cite{autodesk_inc_fusion_2023,freecad_python_2023}.
These applications facilitate the re-use of additional information from the model within the model.
However, programming-based CAD makes this task more challenging.
Applications such as OpenSCAD or JSCAD often limit interactions within the view, preventing users from selecting specific parts or re-using information, for example, the position of an object. Applications could use visual elements as rulers that allow for measurement or even retrieve raw positions, orientations, or sizes from objects directly from the view when clicking on parts of the model's visual representation.
Implementing bidirectional programming \cite{mcguffin_categories_2020}  behaviors could support this task as has been done in SVG \cite{hempel_sketch-n-sketch_2019} and CAD previous work \cite{keeter_antimony_2023, gonzalez_introducing_2023}.

In general, programming-based CAD could benefit from allowing a more enriched interaction in the derived information rendered in the view while coherently connecting it with the code and supporting the edit based on this information.
Moving towards bidirectional programming \cite{hempel_sketch-n-sketch_2019} or live programming \cite{victor_bret_2013} paradigms could bridge the gap existing between both spaces, code and view, which is one significant difficulty in programming \cite{norman_cognitive_1986}.

\subsection{3D printing challenges.}

The interviews revealed problems in the process related to the disconnection between the CAD model and the target environment as described in section \ref{sec::theme3_fabrication}.

\paragraph{Fit between the design object and other physical objects}
Users cannot consider contextual limitations, resulting in longer processes and creativity limitations \cite{shneiderman_creativity_2006}.
P9 explained that the lack of context resulted in more iterations fabricating a case for an emergency button.
Bridging the digital design and the physical environment can facilitate the design process.
One approach is incorporating digital references of the physical environment into the digital design.
Some participants upload STL replicas of objects into OpenSCAD and FreeCAD to have a reference of the object and design around it.
Websites such as Thingiverse~\cite{thingiversecom_thingiverse_2022} or MyMiniFactory~\cite{myminifactory_myminifactory_2022} offer models, mostly in STL format, that users can use as references when designing, although participants and previous work have reported some problems with the search engines and meshes quality \cite{liang_customizar_2022,champion_survey_2020}.
Another possibility is to bring the model into the environment.
DesignAR \cite{reipschlager_designar_2019}, for instance, allows designers to work in augmented reality environments and place models in physical environments using direct manipulation.
Programming-based CAD applications could explore having virtual or augmented reality previews for a realistic preview of objects.
Moreover, these environments could also facilitate code interaction and understanding, as previously explored in other fields \cite{khaloo_code_2017,schutz_live_2019, hori_codehouse_2019, segura_castillo_vr-based_2021}.
Furthermore, expanding interaction in programming-based CAD can leverage bidirectional programming~\cite{mcguffin_categories_2020, hempel_sketch-n-sketch_2019, gonzalez_introducing_2023}.

\paragraph{Include physical measurements}
Capturing data from physical objects has been also reported as challenging  \cite{kim_understanding_2017,ramakers_measurement_2023,mahapatra_barriers_2019}.
Participants reported difficulties in measuring curved and organic shapes using programming-based CAD.
Research could explore sensing devices to capture and transfer information to CAD applications.
Some participants use photogrammetry and scanners to capture the outline of curves and reproduce them digitally, but these solutions were found to be time-consuming, complex, and imprecise.
Additionally, retrieved information as a point cloud is difficult to parameterize.
An alternative solution is to use Bezier curves, where a few control points mathematically define a contour.
Solutions like ShArc \cite{shahmiri_sharc_2020} can capture data from these control points with less effort, creating a suitable solution to replicate organic shapes parametrically.
The use of augmented reality could also help to capture control points to create Bezier curves.

\paragraph{Other challenges in 3D printing}
Users go through different disconnected stages when 3D printing.
CAD applications often limit functionalities to the design, ignoring limitations concerning the printing process.
For example, participants used to apply spatial transformations to their models to locate them in an optimal position and orientation for 3D printing.
However, if any edit was required, they removed these transformations to see the model in a more familiar position and orientation.
Similarly, when participants needed to split a model into parts.
They first finish the model and then apply a \texttt{difference} and \texttt{intersection} with conditionals to save two different STL files.

CAD application could facilitate the design by including information related to the printing process.
For example, the same model would not print the same on different printers or with different materials.
One of the main factors of uncertainty expressed by the participants was the tolerances and clearances.
After printing and detecting problems, they would have to go to the CAD model, adjust it, export it, and print it again.
CAD software could inform users of possible problems related to the design complexity (\textit{e.g.} need for supports), printer tolerances, printer capabilities (\textit{e.g.} possible angles of printing), or materials properties when printing.

\section{Limitations}

The hands-on exercise was a short observation task rather than a controlled user study.
Findings related to it may be limited, missing other challenges that users face in the design process.
Moreover, having experience in direct manipulation programs was not an exclusion criterion.
Consequently, some of the participants had no previous experience in such software, and their answers related to these applications were not based on a reasonable experience and understanding of the direct manipulation paradigm.
Finally, our interview only included OpenSCAD users.
Although most programming-based CAD applications also use a CSG representation, each tool provides different features, and not all of our findings may be generalizable.
Further, a few programming-based tools that use B-rep representation, such as CadQuery, may provide a different user experience and challenges than the ones reported in our findings.

\section{Conclusion}

We interviewed twenty users of the most popular programming-based CAD tool, OpenSCAD, to investigate their motivations and challenges in the design of 3D objects and the 3D printing process.
During these interviews, we included hands-on experience to observe behaviors and difficulties when navigating the model.
With the information collected, we performed a reflexive thematic analysis in an iterative process, developing main themes related to the user's profile, design experience, and printing experience.
Our findings reveal that users are motivated to use programming-based CAD tools thanks to their parametric capability, the possibility of using mathematical expressions, and the precision for 3D printing.
Moreover, it reveals several challenges in connecting the code with the view, understanding and performing spatial transformations, measuring and designing organic and curve shapes, validating dimensions in the view, and re-using pre-existing models.
Programming-based CAD could facilitate some of these tasks by enabling the information that the system stores and effectively communicating it to the user.
Last, our findings also reveal difficulties in the 3D printing process, such as handling uncertainty introduced by printers and material properties, identifying code locations to perform correction based on physical inspection, and validation before printing.

\bibliographystyle{ACM-Reference-Format}
\bibliography{references.bib}

\appendix

\section*{APPENDIX}\label{sec::appendix}
\setcounter{section}{1}

Base questionnaire used in the semi-structured interviews.

\begin{enumerate}
\item  Are you at least an Advanced Beginner with OpenSCAD? (Are you capable of creating designs and understanding the code of a model?)

\item 	What is your gender?

\item 	How old are you?

\item 	What is your academic background?

\item 	What is your current job?

\item 	Do you have experience with 3D printing?

\item 	If yes, tell me about your experience with 3D printing.	When did you start? What were your motivations?

\item 	How often do you 3D print?
\end{enumerate}
\begin{itemize}
\item Daily
\item Weekly
\item Monthly
\item Every two months
\item Every semester
\item Less often
\end{itemize}

Here, we define the terms of direct manipulation and programming-based paradigms that we use in the rest of the interview.

\begin{enumerate}
\setcounter{enumi}{8}
\item  What \textbf{direct manipulation CAD} applications have you used before, and what is your experience in each? Name of the application and skill level
\end{enumerate}

\begin{itemize}
\item 1 - Novice
\item 2 - Advanced Beginner
\item 3 - Competent
\item 4 – Proficient
\item 5 - Expert
\end{itemize}

\begin{enumerate}
\setcounter{enumi}{9}
\item Other than CAD, what other \textbf{programming languages}, in general, have you used, and what is your skill level in each one? Programming language name and skill level
\end{enumerate}
\begin{itemize}
\item 1 - Novice
\item 2 - Advanced Beginner
\item 3 - Competent
\item 4 – Proficient
\item 5 - Expert
\end{itemize}

\begin{enumerate}
\setcounter{enumi}{10}
\item What is your skill level in \textbf{OpenSCAD} ?
\end{enumerate}

\begin{itemize}
\item 1 - Novice
\item 2 - Advanced Beginner
\item 3 - Competent
\item 4 – Proficient
\item 5 - Expert
\end{itemize}

\begin{enumerate}
\setcounter{enumi}{11}
\item  What motivated you to learn/use OpenSCAD specifically?
How did you start?
Did you try other applications?
Why OpenSCAD and no others?

\item 	Let's talk about the last three objects you 3D printed. Describe the object and motivation.

\item 	Would you say that, in general, you 3D print for the motivations mentioned before, or are there other main reasons you print for?

\item 	How did you get the design for those objects?
Design them from scratch,
Pre-existing models

\item 	How do you normally get your models?
Design them from scratch,
Pre-existing models

\item 	What CAD applications did you use to design/edit your last three objects and why?

\item 	What were the major difficulties you found in the process of fabricating these objects (Including all the processes, ideation, design, configuration, printing, iteration, etc)?
What is the most time-consuming part?
What brings more uncertainty? (What makes you iterate more?)

\item 	In general, what are the most difficult parts of the fabrication process (including all the processes, ideation, design, configuration, printing, iteration, etc)?
What is the most time-consuming part?
What brings more uncertainty? (What makes you iterate more?)

\item 	What factors bring more uncertainty or usually make you iterate more times?

\item 	Specifically in the model design part, what is the most difficult and time-consuming part?
Is it something related to the software?
Is it different when you use direct manipulation than programming-based?

\item If different than the previous answer, Specifically in OpenSCAD, what is the most difficult and time-consuming part?

\item 	Do you need to measure physical sizes to transfer them into the digital design?
How do you do it?
What tools and strategies do you use?
How do you verify the correctness of the measurements?

\item 	Tell me about measurement difficulties,
Linear measurements,
Curved and organic shape measurements

\item 	Generally, when you 3D print, how often do you design the models you print from scratch? Why?
\end{enumerate}

\begin{itemize}
\item (0\%) Never
\item (1\% - 20\%) Rarely
\item (20\% - 40\%) Often
\item (40\% - 60\%) Sometimes
\item (60\% - 80\%) Frequently
\item (80\% - 99\%) Very frequently
\item (100\%) Always
\end{itemize}

\begin{enumerate}
\setcounter{enumi}{25}
\item  Generally, when you 3D print, how often do you use a pre-existing model for the models you print (previous projects, friend's model, website model)? Why?
\end{enumerate}

\begin{itemize}
\item (0\%) Never
\item (1\% - 20\%) Rarely
\item (20\% - 40\%) Often
\item (40\% - 60\%) Sometimes
\item (60\% - 80\%) Frequently
\item (80\% - 99\%) Very frequently
\item (100\%) Always
\end{itemize}

\begin{enumerate}
\setcounter{enumi}{26}
\item 	What motivates you to design from scratch or to use a pre-existing model?

\item 	Do you know model-storing websites such as Thingiverse? What others?

\item 	If yes, what do you think about them? Do you use them? Do you find them useful?

\item 	If you know Thingiverse, have you used the Customizer tool? Talk to me about your experience with this tool.

\item 	When you use pre-existing models, in what format do you get them? (stl, obj, code...)

\item 	When you re-use a \textbf{non-coded }pre-existing model, how often do you need to edit it? What types of modifications do you make?
\end{enumerate}

\begin{itemize}
\item (0\%) Never
\item (1\% - 20\%) Rarely
\item (20\% - 40\%) Often
\item (40\% - 60\%) Sometimes
\item (60\% - 80\%) Frequently
\item (80\% - 99\%) Very frequently
\item (100\%) Always
\end{itemize}

\begin{enumerate}
\setcounter{enumi}{32}
\item  When you re-use a \textbf{non-coded} pre-existing model, how difficult is it to edit it? Explain what applications you use and why the level of difficulty you selected.
\end{enumerate}

\begin{itemize}
\item 1- Very easy
\item 2- Easy
\item 3- Neutral
\item 4- Difficult
\item 5- Very difficult
\end{itemize}

\begin{enumerate}
\setcounter{enumi}{33}
\item  When you re-use a \textbf{coded} pre-existing model, how often do you need to edit it? What types of modifications do you make?
\end{enumerate}

\begin{itemize}
\item (0\%) Never
\item (1\% - 20\%) Rarely
\item (20\% - 40\%) Often
\item (40\% - 60\%) Sometimes
\item (60\% - 80\%) Frequently
\item (80\% - 99\%) Very frequently
\item (100\%) Always
\end{itemize}

\begin{enumerate}
\setcounter{enumi}{34}
\item  When you re-use a \textbf{coded} pre-existing model, how difficult is it to edit it? Explain what applications you use and why the level of difficulty you selected.
\end{enumerate}

\begin{itemize}
\item 1- Very easy
\item 2- Easy
\item 3- Neutral
\item 4- Difficult
\item 5- Very difficult
\end{itemize}

\begin{enumerate}
\setcounter{enumi}{35}
\item 	Did you bring some of your previous OpenSCAD projects?
Talk to me about one of them,
How was the process, how many iterations did you need, and what was the most difficult part of the process?

\item 	\textbf{(Hands-on exercise)} I will ask you to localize in the code the specific statements that create a part that I will point out in the view. Share aloud the thinking process you follow to find it. Is this a task you normally do when designing: looking for a specific part in the code based on the view? What is the hardest part of doing it? What strategies do you use normally?

\item 	How difficult was the task?

\item 	In OpenSCAD (and programming-based), how easily can you link the output in the view to the code?

\item 	What would you say is the best of OpenSCAD and the worst? What would you say is the best of programming-based CAD and the worst?

\item 	What are the advantages and disadvantages of direct manipulation and programming-based applications like OpenSCAD? When do you prefer to use one or the other?

\end{enumerate}

\end{document}